\documentclass[twoside,leqno,twocolumn,english]{article}
\usepackage{ltexpprt}

\usepackage{times}
\usepackage{amsmath}
\usepackage{graphicx}
\usepackage{balance}
\usepackage{algorithm}
\usepackage{algorithmic}
\usepackage{amssymb}
\usepackage{amsfonts}
\usepackage{url}
\usepackage{multirow}
\usepackage{color}
\usepackage{bm}

\newcommand{\hide}[1]{}

\newcommand{\LINEIFONE}[2]{\STATE \algorithmicif\ {#1}\ \algorithmicthen\ {#2} \algorithmicend\ \algorithmicif}
\newcommand{\LINEIF}[2]{\STATE \algorithmicif\ {#1}\ \algorithmicthen\ {#2} }

\newcommand{\rls}{\textsc{RLS-Problem}}

\newcommand{\seedrls}{\textsc{Seeded-RLS-Problem}}
\newcommand{\krls}{\textsc{$k$RLS-Problem}}
\newcommand{\rgs}{\textsc{RGS-Problem}}

\newcommand{\greedymrs}{\textsc{GreedyRLS}}
\newcommand{\grasp}{\textsc{GRASP}}
\newcommand{\gmrs}{\textsc{GRASP-RLS}}

\newcommand{\cons}{\textsc{GRASP-RLS-Construction}}
\newcommand{\local}{\textsc{GRASP-RLS-LocalSearch}}

\newtheorem{problem}{Problem}
\newtheorem{definition}{Definition}

\newfont{\secpar}{ptmb8t at 10pt}

\newcommand{\A}{$\mathbf{A}$}

\newcommand{\uj}{$\mathbf{u_j}$}
\newcommand{\mat}{\mathbf}
\newcommand{\bit}{\begin{itemize}}
\newcommand{\eit}{\end{itemize}}
\newcommand{\ben}{\begin{enumerate}}
\newcommand{\een}{\end{enumerate}}
\newcommand{\beq}{\begin{equation}}
\newcommand{\eeq}{\end{equation}}

\title{Where Graph Topology Matters: The Robust Subgraph Problem}

\begin{document}
\author{
\makebox[230pt]{Hau Chan \quad \quad \quad Shuchu Han \quad \quad \quad  Leman Akoglu}\\
\makebox[230pt]{Stony Brook University}\\
\makebox[230pt]{\{hauchan, shhan, leman\}@cs.stonybrook.edu} \\
\vspace{-0.25in}
}
\date{}
\maketitle

\begin{abstract}

Robustness is a critical measure of the resilience of large networked systems, such as 
transportation and communication networks. 
Most prior works focus on the global robustness of a given graph at large, e.g., by measuring its overall vulnerability to external attacks or random failures. 
In this paper, we turn attention to local robustness and pose a novel problem in the lines of subgraph mining: given a large graph, how can we find its most robust local subgraph ({\sc RLS})? 

We define a robust subgraph as a subset of nodes with high communicability \cite{journals/corr/abs-1109-2950} among them, and formulate the \rls~of finding a subgraph of given size with maximum robustness in the host graph. 
Our formulation is related to the recently proposed general framework \cite{conf/kdd/TsourakakisBGGT13} for the densest subgraph problem, however differs from it substantially in that besides the number of edges in the subgraph, robustness also concerns with the \textit{placement} of edges, i.e., the subgraph topology.
We show that the \rls~is {\bf NP}-hard and propose two heuristic algorithms based on top-down and bottom-up search strategies. Further, we present modifications of our algorithms to handle three practical variants of the  \rls.
Experiments on synthetic and real-world graphs demonstrate that we find subgraphs with larger robustness than the densest subgraphs \cite{conf/approx/Charikar00,conf/kdd/TsourakakisBGGT13} even at lower densities, suggesting that the existing approaches are not suitable for the new problem setting.   

\end{abstract}


\vspace{-0.06in}
\section{Introduction}
\label{sec:intro}

Complex networked systems, such as the Internet, road networks, communication networks, the power grid, etc., are a major part of our modern world.
The performance and reliable functioning of complex networks depend on their structural robustness, 
e.g., their ability to retain functionality in the face of damage to parts of the network \cite{Watts2003}. 

Robustness has been studied in many fields  including physics, biology, mathematics, and networking.
The research areas include 
quantifying robustness of a network \cite{journals/corr/EllensK13,Holme2002,MalliarosMF12,WUJun78902},  
studying the response of networks to various attack strategies \cite{albert2000error,callaway2000nra,citeulike2352886,expansion2006,Holme2002,conf/cikm/TongPEFF12},
manipulating a network to improve its overall robustness \cite{Beygelzimer2005,ChanSDM14,journals/amc/SydneySG13,journals/corr/abs-1203-2982}, and
 designing optimally robust networks from scratch \cite{FrankFrisch1970,journals/telsys/GhamryE12,RePEc,Shargel03}.


A vast majority of prior work has focused on the global robustness of graphs at large.
On the other hand, research on local robustness is limited to a few works, e.g., on finding robust subgraphs with large spectral radius \cite{journals/jucs/AndersenC07} and identifying critical regions  \cite{conf/infocom/TrajanovskiKM13}.
In this paper, we turn attention to local robustness and pose a novel subgraph mining problem: given a large graph, how can we find its most robust local subgraph of a given size?

Our measure of robustness is the natural connectivity which is based on the reachability of the nodes, also phrased as their ``communicability'' \cite{WUJun78902}. 
As we introduced in prior work \cite{ChanSDM14}, it exhibits several
desirable properties; e.g., it captures redundancy by quantifying the count and length of 
alternative/back-up paths between the nodes.
As such, robust subgraphs are intuitively sets of nodes with high communicability among each other.
From the practical point of view, they may form the cores of larger communities or constitute the central backbones in large networks, maintaining connectivity and communication at large  \cite{journals/corr/abs-1109-2950}. 




While the robust subgraph problem has not been studied before, 
similar problems have been addressed (\S\ref{sec:related}). 
Probably the most similar to ours is the densest subgraph problem, aiming to find subgraphs with highest
average degree  \cite{journals/jal/AsahiroITT00,conf/approx/Charikar00,Goldberg84} or edge density \cite{journals/dam/PattilloVBB13,conf/kdd/TsourakakisBGGT13}. 
However, density is different from robustness; while the former
concerns with the number of edges in the subgraph, the topology is also of concern for the latter (\S\ref{subsec:denserobust}).
We offer the following contributions.

\vspace{-0.1in}
\bit
\setlength{\itemsep}{-1.0\itemsep}
\item We formulate a new problem of
finding the most robust local subgraph ({\sc RLS}) in a given graph. While in the line of subgraph mining problems, it has not been studied  theoretically before (\S \ref{ssec:define}).

\item 
We show that \rls~is {\bf NP}-hard, and further study its heredity and monotonicity properties (\S \ref{ssec:properties}).

\item We propose two fast heuristic algorithms to solve the \rls~for large graphs: a top-down greedy algorithm that iteratively removes nodes, and a bottom-up approach based on the greedy randomized adaptive search procedure (GRASP) \cite{Feo95greedyrandomized} (\S \ref{sec:algos}).

\item We introduce three practical variants of the \rls~(\S \ref{sub:var}); 
and show how to modify our  algorithms to address these problem variants (\S \ref{sec:algos}).


\eit
\vspace{-0.1in}

We extensively evaluate our methods on both synthetic and real-world graphs.
As our \rls~is a new one, we compare to three algorithms (one in \cite{conf/approx/Charikar00}, two in \cite{conf/kdd/TsourakakisBGGT13}) for the densest subgraph problem.
We find subgraphs with higher robustness than the densest subgraphs even at lower densities, demonstrating that 
the existing algorithms are not applicable for the new problem setting (\S \ref{sec:experiments}).

\vspace{-0.1in}
\section{Background and Preliminaries}
\label{sec:prelim}


\subsection{Graph Robustness}
\label{subsec:robst}

Robustness is a critical property of large-scale networks, and thus has been studied in physics, mathematics, computer science, and biology. As a result, there exists a diverse set of robustness measures, e.g., mean shortest paths, efficiency, pairwise connectivity, etc. \cite{journals/corr/EllensK13}.  

In this paper, we adopt a spectral measure of robustness called {\em natural connectivity}  \cite{WUJun78902}, written as 

\vspace{-0.2in}
\beq
\label{natur}
\bar{\lambda}(G) = \log (\frac{1}{n} \sum_{i=1}^n e^{\lambda_i})\;, 
\eeq
\vspace{-0.15in}

\noindent
which can be thought of as the ``average eigenvalue'' of graph $G$, where
$\lambda _1 \geq \lambda _2 \geq ... \geq \lambda _n$ denote a non-increasing ordering of the eigenvalues of its adjacency matrix $\mathbf{A}$.

Among other desirable properties \cite{ChanSDM14}, natural connectivity
is interpretable; it is directly related to the subgraph centralities ($SC$) in the graph.
The $SC(i)$ of a node $i$ is known as its communicability \cite{journals/corr/abs-1109-2950}, and
is based on the ``weighted'' sum of the number of closed walks that it participates in:

\vspace{-0.2in}
\begin{align}
S(G) &= \sum_{i=1}^n SC(i)=\sum_{i=1}^{n} \sum_{k=0}^{\infty} \frac{(\mathbf{A}^k)_{ii}} {k!}\;, \nonumber 
\vspace{-0.4in}
\end{align}
\vspace{-0.1in}

\noindent
where $(A^k)_{ii}$ is
the number of closed walks of length $k$ of node $i$. The $k!$ scaling
ensures that the weighted sum does not diverge, and longer walks count less.
$S(G)$ is also referred to as the Estrada index \cite{journals/corr/abs-1109-2950} which strongly correlates with the folding degree of proteins \cite{journals/bioinformatics/Estrada02}. 

Noting that $\sum_{i=1}^{n} (\mathbf{A}^k)_{ii} = \text{trace}(\mathbf{A}^k) = \sum_{i=1}^{n} \lambda_i^k$ and by Taylor series of the exponential function we can write
\begin{align}
\vspace{-0.2in}
S(G) &= \sum_{k=0}^{\infty} \sum_{i=1}^{n} \frac{(\mathbf{A}^k)_{ii}} {k!} = 
 \sum_{i=1}^{n} \sum_{k=0}^{\infty} \frac{\lambda_i^k} {k!} = \sum_{i=1}^{n} e^{\lambda_i} \;.\nonumber 
\vspace{-0.3in}
\end{align}

As such, natural connectivity is the normalized Estrada index and quantifies the ``average communicability'' in $G$. 

\vspace{-0.1in}
\subsection{Robustness vs.\ Density}
\label{subsec:denserobust}

Graph robustness appears to be related to graph density; however as we show here, there exist key distinctions between them. 

\begin{figure}[!b]
\vspace{-0.2in}
\centering
\begin{tabular}{cp{0.025in}cp{0.025in}c}
\includegraphics[width=0.7in]{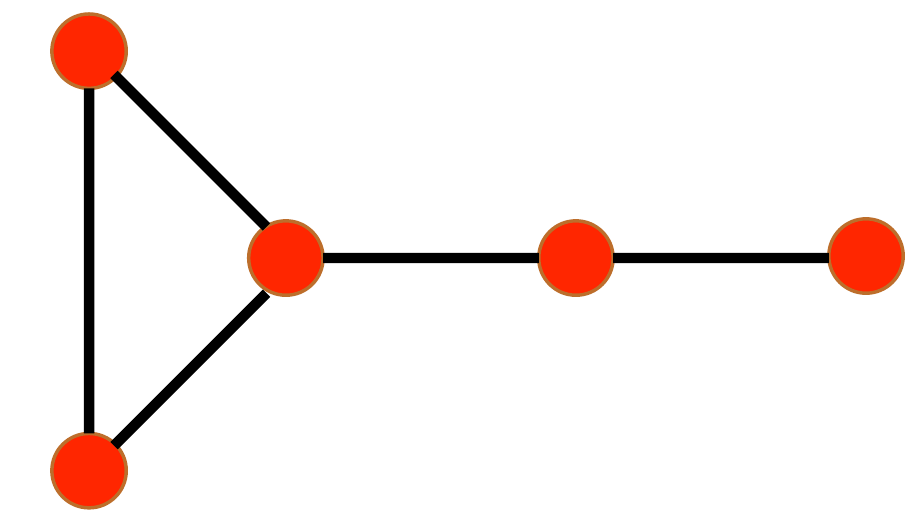}  & 
& \includegraphics[width=0.9in]{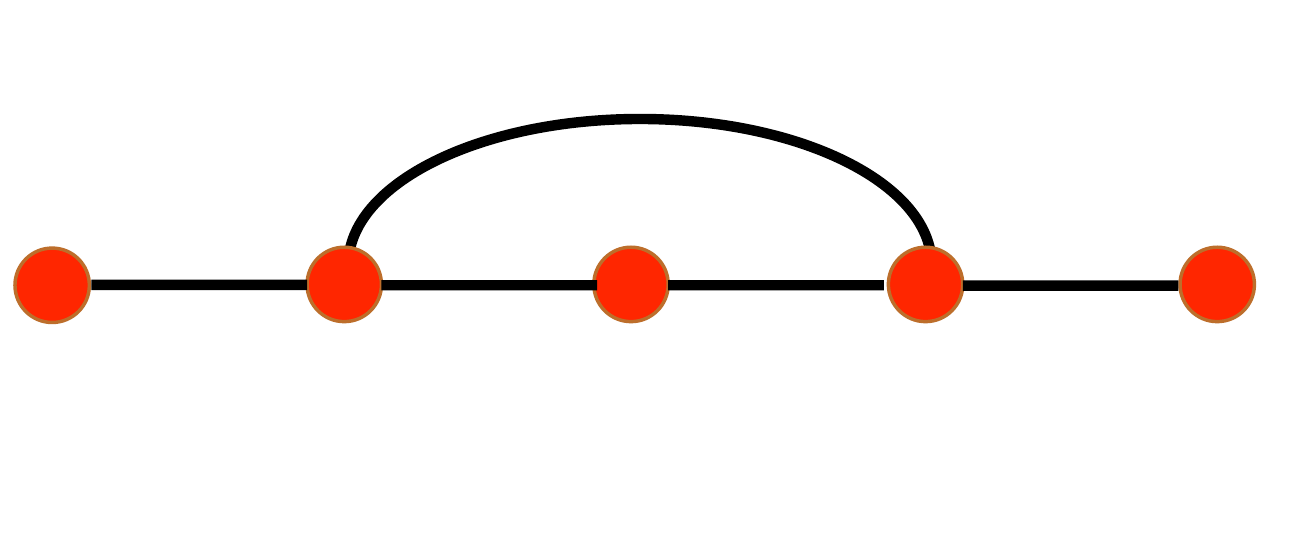} & 
&\includegraphics[width=0.4in]{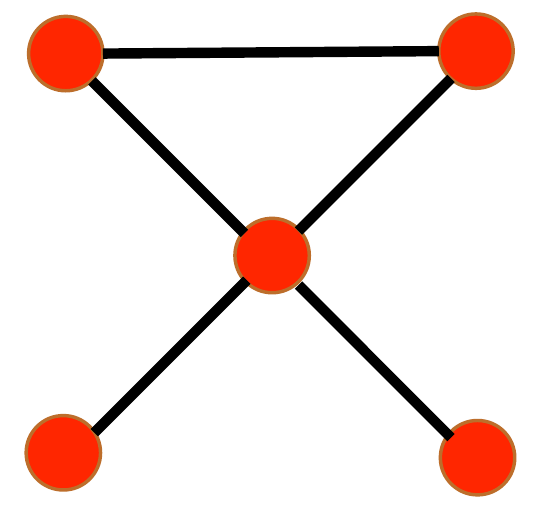} \\
	$\bar{\lambda}$ = 0.9564 && $\bar{\lambda}$ = 0.9804 && $\bar{\lambda}$ = 0.9965 \\
\end{tabular}
\vspace{-0.15in}
\caption{\small Example graphs with the same density but different robustness, due to their distinct graph topology.}
\label{fig:diffrobust}
\vspace{-0.15in}
\end{figure}

Firstly, while density 
directly uses the number of edges $e$, such as $\frac{2e(G)}{|V|}$ as in average degree \cite{journals/jal/AsahiroITT00,conf/approx/Charikar00,Goldberg84} or $\frac{2e(G)}{|V|(|V|-1)}$ as in edge density  \cite{journals/dam/PattilloVBB13,conf/kdd/TsourakakisBGGT13}, robustness follows an indirect route; it quantifies the count and length of paths and uses the graph spectrum. Thus, the objectives of robust and dense subgraph mining problems are distinct.

More notably, density concerns with the number of edges in the graph and not with the topology. On the other hand, for robustness the {\em placement} of edges (i.e., topology) is as much, if not more important. 
In fact, graphs with the same number of nodes and edges but different topologies are indistinguishable from the density point of view (Figure \ref{fig:diffrobust}).

\begin{figure}[t]
\vspace{-0.05in}
\centering
\begin{tabular}{c}
\includegraphics[width=2.0in]{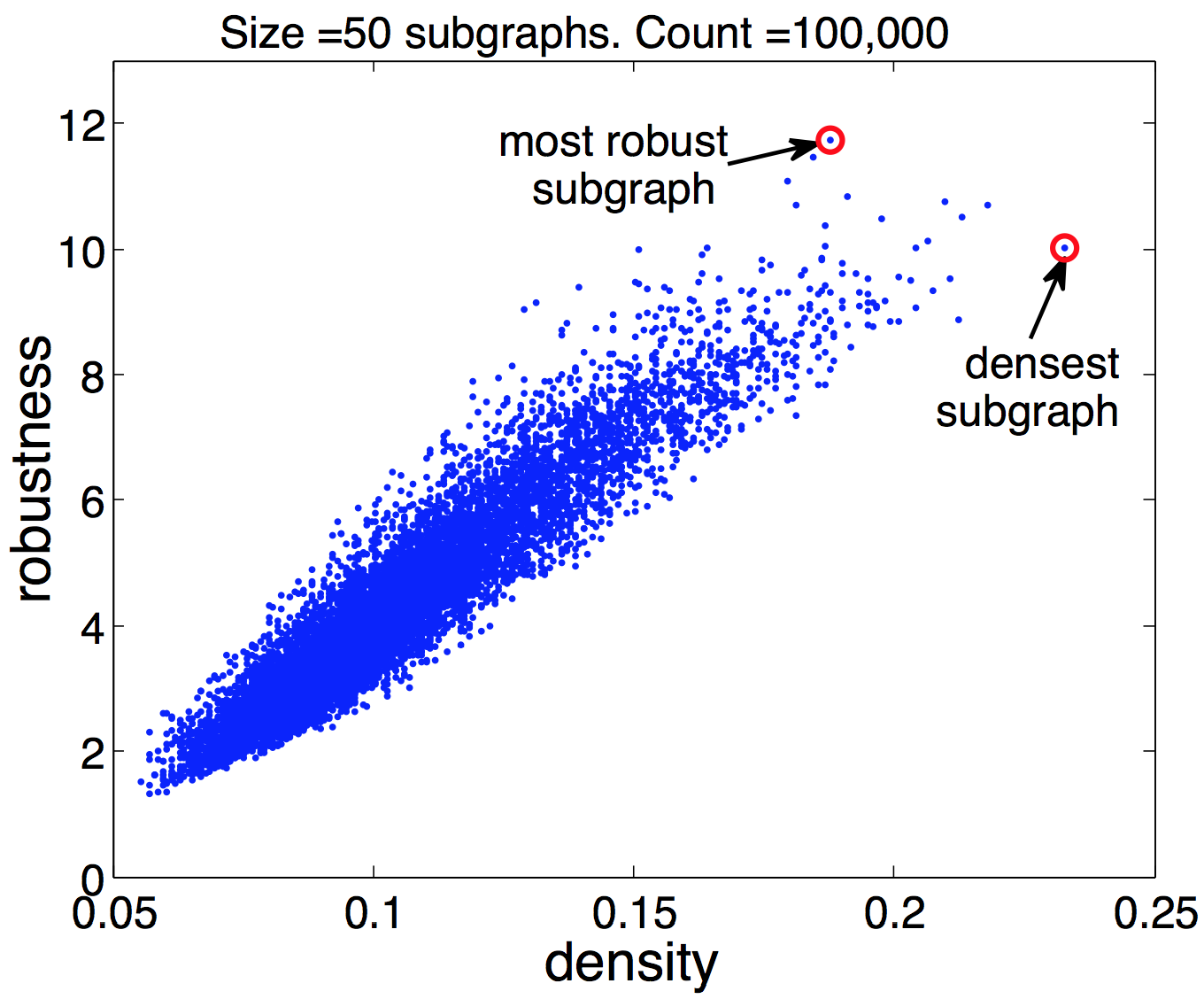}\\
\end{tabular}
\vspace{-0.15in}
\caption{\small Robustness vs. Density of 100,000 connected subgraphs (blue dots) from a real-world email network.}
\label{fig:diftopo}
\vspace{-0.15in}
\end{figure}

To illustrate further, we show in Figure \ref{fig:diftopo} the robustness vs. density of example subgraphs, each of size $50$, sampled\footnote{\scriptsize We create subgraphs by snowball sampling: pick a random node  and progressively add its neighbors with probability $p$, and iterate in a depth-first fashion. Connectivity is guaranteed by adding at least one neighbor of each node. We use varying $p \in (0,1)$ to control the tree-likeness, and obtain subgraphs with various densities.}
from a real-world email network (\S\ref{sec:experiments}, Table \ref{tab:data}). While the two properties are correlated, subgraphs with the same density can have a range of different robustness. In fact, among the samples, the densest and the most robust subgraphs are distinct, indicating that one does not always imply the other.

\vspace{-0.1in}

\section{Robust Local Subgraphs}
\label{sec:proposed}

\subsection{Problem Definition}
\label{ssec:define}


In their inspiring work \cite{conf/kdd/TsourakakisBGGT13}, 
Tsourakakis \textit{et al.} recently defined a 
general framework for subgraph density functions, which
is written as

\vspace{-0.15in}
\begin{equation*}
f_{\alpha} (S) = g(e[S]) - \alpha h(|S|)\;,
\end{equation*}
where $S \subseteq V$ is a set of nodes, $S \neq \emptyset$,
$e[S]$ is the number of edges in the subgraph induced by $S$,
$\alpha > 0$, and $g$ and $h$ are any two strictly increasing functions.

Under this framework, maximizing the \textit{average degree} of a subgraph \cite{journals/jal/AsahiroITT00,conf/approx/Charikar00,Goldberg84} corresponds to $g(x) = h(x) = \log x$ and $\alpha=1$ such that

\vspace{-0.15in}
\begin{equation*}
f(S) = \log \frac{e[S]}{|S|} \;\;.
\end{equation*}
\vspace{-0.1in}

In order to define our problem, we can relate the objective of our setting to this general framework.
Specifically, our objective can be written as

\vspace{-0.125in}
\begin{equation*}
f(S) = \log \frac{\sum_{i=1}^{|S|} e^{\lambda_i}}{|S|} \;\;,
\end{equation*}
which is to maximize the \textit{average eigenvalue} of a subgraph.
Therefore, the objectives of the two problems are distinct, although they both fall under a more general framework \cite{conf/kdd/TsourakakisBGGT13}.

In the following we formally define our robust local subgraph mining problem, which is to find the highest robustness subgraph of a certain size (hence the locality) in a given graph, which we call the \rls.

\vspace{-0.05in}
\begin{problem}[\rls]
Given a graph $G = (V,E)$ and an integer $s$, find a subgraph with nodes $S^* \subseteq V$ of size $|S^*|=s$ such that
\vspace{-0.125in}
$$
f(S^*) = \log \sum_{i=1}^{s} e^{\lambda_i([S^*])} - \log s \geq f(S), \;\; \forall S \subseteq V, |S|=s.
$$
\vspace{-0.15in}

$S^*$  is referred as the \textsf{\em most robust $s$-subgraph}.
\end{problem}
\vspace{-0.075in}

One can interpret a robust subgraph as containing a set of nodes having large communicability within the subgraph.


\vspace{-0.1in}
\begin{theorem}
\label{th:hardness}
The optimal \rls~is {\bf NP}-Hard.
\end{theorem}

\vspace{-0.2in}
\begin{proof}
See Appendix \ref{sec:npproof}.
\end{proof}
\vspace{-0.05in}


\vspace{-0.15in}
\subsection{Problem Properties}
\label{ssec:properties}
\sloppy{
Certain characteristics of hard combinatorial problems sometimes guide the development of approximation algorithms for those problems. 
In this work, we study two such characteristics, namely semi-heredity and subgraph monotonicity, for the \rls.
}

Problems that exhibit the (semi-)heredity or monotonicity properties often enjoy algorithms that explore the search space in a smart and efficient way.
For example cliques exhibit heredity, i.e., all induced subgraphs are also cliques.
This is a key property used in successful algorithms for the maximum clique problem, e.g., checking maximality by inclusion is a trivial task and effective pruning strategies can be employed within a branch-and-bound framework. Other algorithms exploit monotonicity to employ ``smart node ordering'' strategies to find iteratively improving solutions. Such orderings help starting with a promising node and sequentially adding the next node in the order such that the resulting subgraphs all satisfy some desired criteria, like a minimum density, which enables finding large solutions quickly.

\vspace{-0.1in}
\begin{theorem}
\label{th:semiheredity}
Robustness $\bar{\lambda}$ is not semi-hereditary. That is, a graph with $\bar{\lambda}=\alpha$ and $s>1$ nodes is not always a strict superset of {\em some} graph with $s-1$ nodes and $\bar{\lambda}\geq \alpha$.
\end{theorem}

\vspace{-0.2in}
\begin{theorem}
\label{th:monoton}
Robustness $\bar{\lambda}$ is not subgraph monotonic.
\end{theorem}

\vspace{-0.2in}
\begin{proof}
See Appendix \ref{sec:properties} for definitions and proofs.
\end{proof}
\vspace{-0.05in}

Alas, robust subgraphs do not exhibit any of these properties. This suggests that our \rls~is likely harder than the maximum clique and densest subgraph problems as, unlike robust subgraphs, (quasi-)cliques are shown to exhibit e.g., the (semi-)heredity property \cite{journals/dam/PattilloVBB13}.

\vspace{-0.1in}
\subsection{Problem Variants}
\label{sub:var}

In Appendix \ref{sec:variants}, we introduce three practical variants of our \rls: finding ($i$) the most robust subgraph (no size constraint), ($ii$) top-$k$ most robust $s$-subgraphs, and ($iii$) the most robust $s$-subgraph including a given set of seed nodes. 
We also show how to adapt our algorithms for the \rls~to these variants (\S\ref{sec:algos}).

\section{Finding Robust Local Subgraphs}
\label{sec:algos}


Given the hardness of the \rls, we design two heuristic solutions.
The first is \greedymrs, a top-down approach that iteratively removes nodes to obtain a subgraph of desired size.
This greedy strategy serves as a simple baseline.
Our second and proposed solution \gmrs~is a bottom-up randomized approach in which we iteratively add nodes to build up our subgraphs.
Both solutions order the nodes by their contributions to the robustness.

\vspace{-0.1in}
\subsection{Greedy Top-down Search Approach}
\label{subsec:greedy}

This approach iteratively and greedily removes the nodes one by one from the given graph $G=(V,E)$, $|V|=n, |E|=m$, until a subgraph with the desired size $s$ is reached. At each iteration, the
node whose removal results in the maximum robustness of the residual graph is selected for removal.\footnote{\scriptsize Robustness of the residual graph can be lower or higher; $S(G)$ decreases due to monotonicity, but the denominator also shrinks to $(s-1)$ at every step.}

The removal of a node involves removing the node itself and the edges attached to it from the graph, where the residual graph becomes $G[V\backslash \{i\}]$. Let $i$ denote a node to be removed.
Let us then write the updated  robustness $\bar{\lambda}_\Delta$ as

\vspace{-0.25in}
\beq
\label{update}
\bar{\lambda}_\Delta = \log \bigg(\frac{1}{n-1} \sum_{j=1}^{n-1} e^{\lambda_j + \Delta \lambda_j}\bigg)\;.
\eeq
\vspace{-0.2in}

As such, we are interested in identifying the node that
maximizes $\bar{\lambda}_\Delta$, or equivalently

\vspace{-0.25in}
\begin{align}
\label{nodeselect}
{\bm \max.} \hspace{0.1in} & e^{\lambda_1 + \Delta \lambda_1} + e^{\lambda_2 + \Delta \lambda_2} + \ldots + e^{\lambda_{n-1} + \Delta \lambda_{n-1}}  \\ 
 & e^{\lambda_1} (e^{\Delta \lambda_1} + e^{(\lambda_2-\lambda_1)} e^{\Delta \lambda_2} + \ldots + e^{(\lambda_{n-1}-\lambda_1)} e^{\Delta \lambda_{n-1}} )  \nonumber  \\
 & e^{\lambda_1} (e^{\Delta \lambda_1} + c_2 e^{\Delta \lambda_2} + \ldots + c_{n-1} e^{\Delta \lambda_{n-1}} )\nonumber
\end{align}
\vspace{-0.2in}

where $c_j$'s denote $e^{\lambda_j - \lambda_1}$ $\forall j \geq 2$ and $c_j \leq 1$.

\vspace{-0.1in}
\subsubsection{Updating the eigen-pairs}
When a node is removed from the graph, its spectrum (i.e., the eigen-pairs $(\lambda_j, \mathbf{u_j})$) also changes.
Recomputing the eigen-values to compute robustness $\bar{\lambda}_\Delta$ every time a node is removed is computationally challenging.
Therefore, we employ fast update schemes based on the first order matrix 
perturbation theory \cite{MatrixPerturb}.

Let $\Delta$\A~ and $(\Delta \lambda_j, \Delta \mathbf{u_j})$ denote the change in \A~and $(\lambda_j, \mathbf{u_j})$  $\forall j$, respectively, where $\Delta$\A~is symmetric.
Suppose after the adjustment \A~becomes
$$
\tilde{\mathbf{A}} = \mathbf{A} + \Delta \mathbf{A}
$$
where each eigen-pair $(\tilde{\lambda}_j, \tilde{u}_j)$  is written as
\vspace{-0.05in}
\begin{align}
\tilde{\lambda}_j  = \lambda_j + \Delta \lambda_j \nonumber \;\;\;\;\;\;  & \text{and} \;\;\;\;\;\;
\mathbf{\tilde{u}_j}  = \mathbf{u_j} + \Delta \mathbf{u_j} \nonumber
\end{align}
\vspace{-0.25in}

\vspace{-0.05in}
\begin{lemma}
\label{lemma:eigenvalueupdate}
Given a perturbation $\Delta$\A~to a matrix \A, its eigenvalues can be updated by 

\vspace{-0.25in}
\beq
\label{updatel}
\Delta \lambda_j = \mathbf{u_j}' \Delta \mathbf{A}  \mathbf{u_j}.
\eeq
\end{lemma}

\vspace{-0.15in}
\begin{proof}
 See Appendix \ref{ssec:eig}.
\end{proof}

Since updating the eigenvalues involves using the eigenvectors, which also change with node removals, we use the following to update the eigenvectors as well.

\vspace{-0.05in}
\begin{lemma}
\label{lemma:eigenvectorupdate}
Given a perturbation $\Delta$\A~to a matrix \A, its eigenvectors can be updated by 

\vspace{-0.2in}
\beq
\label{updateu}
\Delta \mathbf{u_j} =\sum\limits_{i=1, i\neq j}^n \bigg( \frac{\mathbf{u_i}' \Delta \mathbf{A} \mathbf{u_j}}{\lambda_j-\lambda_i} \mathbf{u_i} \bigg).
\eeq
\end{lemma}

\vspace{-0.2in}
\begin{proof}
 See Appendix \ref{ssec:eigv}.
\end{proof}

\vspace{-0.2in}
\subsubsection{Node selection for removal}

By using {\sc Lemma} {\ref{lemma:eigenvalueupdate}},  
we can write the effect of perturbing \A~with the removal of a node $i$ on the eigenvalues as

\vspace{-0.2in}
\beq
\label{picknode1}
\Delta \lambda_j = \mathbf{u_j}' \Delta \mathbf{A}  \mathbf{u_j} = -2 \mathbf{u_{ij}} \sum_{v\in {\cal N}(i)} \mathbf{u_{vj}}
\eeq
\vspace{-0.15in}

\noindent
where $\Delta \mathbf{A}(i,v) = \Delta \mathbf{A}(v,i)= -1$, for $v \in {\cal N}(i)$, and $0$ elsewhere, and ${\cal N}(i)$ denotes the set of neighbors of $i$. 
Thus, at each step we choose the node $i\in V$ that maximizes

\vspace{-0.15in}
\beq
\label{picknode}
e^{\lambda_1} \bigg(e^{-2 \mathbf{u_{i1}} \sum\limits_{v\in {\cal N}(i)} \mathbf{u_{v1}}} + \ldots + c_{n-1} e^{-2 \mathbf{u_{i{n-1}}} \sum\limits_{v\in {\cal N}(i)} \mathbf{u_{v{n-1}}}} \bigg)
\eeq
\vspace{-0.15in}

We remark that it is infeasible to compute all the $n$ eigenvalues of a graph with $n$ nodes, for very large $n$. Thanks to the skewed spectrum of real-world graphs \cite{conf/sigcomm/FaloutsosFF99}, we can rely on the observation that only the top few eigenvalues have large magnitudes. This implies that the $c_j$ terms in Equ. (\ref{nodeselect}) and also Equ. (\ref{picknode}) become much smaller for increasing $j$ and can be ignored. Therefore, we use the top $t$ eigenvalues to approximate the robustness of a graph. In the past, the skewed property of the spectrum has also been exploited to approximate triangle counts in large graphs \cite{conf/icdm/Tsourakakis08}. 

The outline of the \greedymrs~algorithm, 
its complexity analysis, and its adaptations for the \rls~variants (\S\ref{sub:var})
can be found in Appendix \ref{sec:greedyalg}.


\vspace{-0.1in}
\subsection{Greedy Randomized Adaptive Search Procedure (GRASP) Approach}
\label{sec:grasp}

The top-down approach makes a greedy decision at every step.
If the desired subgraphs are small,
however, this incurs many greedy decisions, especially on large graphs where the number of greedy steps $(n-s)$ would be excessive.
Since the \rls~does not exhibit monotonicity or semi-heredity properties (\S\ref{ssec:properties}), taking large number of greedy steps can yield poor performance.
Therefore, we propose a bottom-up approach that performs local operations to build up solutions from scratch. 


Our local approach is based on a meta-heuristic called \grasp~\cite{Feo95greedyrandomized} for solving combinatorial optimization problems.
A \grasp, or greedy randomized adaptive search procedure, is a multi-start or iterative process, in which each iteration consists of two phases: ($i$) a construction phase, in which an initial feasible solution is produced, and ($ii$) a local search phase, in which a better solution with higher objective value
in the neighborhood of the constructed solution is sought. The best overall solution becomes the final result.

The pseudo-code in Algorithm \ref{alg:grasp}
shows the general \grasp~for maximization, where $T_{\max}$ iterations are done.
For maximizing our objective, we use $f : S \rightarrow \mathbb{R} \equiv \bar{\lambda}$, i.e., the robustness function as given in Equ. (\ref{natur}). We next describe the details of our two \grasp~phases.


\vspace{-0.1in}
\begin{algorithm}[h]
\caption{\gmrs} \label{alg:grasp}
\begin{algorithmic}[1]
\REQUIRE Graph $G = (V,E)$, $T_{\max}$, $f(\cdot)$, $g(\cdot)$, integer $s$
\ENSURE Subset of nodes $S^* \subseteq V$, $|S^*|=s$
\STATE $f^* = -\infty$, $S^* = \emptyset$
\FOR{$z = 1, 2, \ldots, T_{\max}$}
\STATE $S \leftarrow$ \cons($G$, $g(\cdot)$, $s$)
\STATE $S' \leftarrow$ \local($G$, $S$, $f(\cdot)$, $s$)
\LINEIF{$f(S') > f^*$} {$S^* \leftarrow S$, $f^* = f(S)$}
\ENDFOR
\RETURN $S^*$
\end{algorithmic}
\end{algorithm}
\vspace{-0.2in}

 \subsubsection{Construction}
In the construction phase, a feasible seed solution is iteratively constructed, one node at a time. At each iteration, the choice of the next node to be added is determined by ordering all candidate nodes $C$  in a restricted candidate list, called $RCL$, with respect to a greedy function 
$g : C \rightarrow \mathbb{R}$, and randomly choosing one of the candidates in the list. 
Candidate set in the first iteration is set to $V$ and in later iterations it contains the nodes in the neighborhood ${\cal N}(S)$ of the current solution $S$.
The size of $RCL$ is determined by a real parameter $\beta \in [0, 1]$, which controls the amount of greediness and randomness. $\beta = 1$ corresponds to a purely greedy construction, while $\beta = 0$ produces a purely random one.  Algorithm \ref{alg:cons} describes our construction phase.

\vspace{-0.05in}
\begin{algorithm}[h]
\caption{\cons} \label{alg:cons}
\begin{algorithmic}[1]
\REQUIRE Graph $G = (V,E)$, $g(\cdot)$, integer $s$
\ENSURE Subset of nodes $S \subseteq V$
\STATE $S \leftarrow \emptyset$, $C \leftarrow V$

\WHILE{$|S| < s$}
\STATE Evaluate $g(v)$ for all $v\in C$
\STATE $\bar{c} \leftarrow \max_{v \in C} g(v)$, $\underline{c} \leftarrow \min_{v \in C} g(v)$
\STATE Select $\beta \in [0, 1]$ using a strategy 
\STATE $RCL \leftarrow \{v\in C| g(v) \geq \underline{c} + \beta(\bar{c}-\underline{c})\}$
\STATE Select a vertex $r$ from $RCL$ at random 
\STATE $S := S \cup \{r\}$, $C \leftarrow {\cal N}(S) \backslash S$
\ENDWHILE
\RETURN $S$
\end{algorithmic}
\end{algorithm}
\vspace{-0.1in}

\noindent
{\bf Selecting $\mathbf{g(\cdot)}$}:
We aim to include locally dense nodes in our seed solutions. Therefore,
in the first iteration of the construction we use
$g(v) = \frac{t(v)}{d(v)}$, where $t(v)$ denotes the number of local triangles of $v$, and $d(v)$ is its degree.
Initially the candidate set $C$ is equal to the node set $V$, thus
we approximate the local triangle counts for speed \cite{conf/icdm/Tsourakakis08}.
In later iterations we use $g(v) = \Delta\bar{\lambda}_v$; the difference in robustness when a candidate node is added to the current subgraph.


\vspace{0.025in}
\noindent
{\bf Selecting $\mathbf{\bm \beta}$}:
Setting $\beta=1$ is purely greedy and produces the same seed subgraph in every \grasp~iteration.
To incorporate randomness while staying close to the greedy best-first selection, we choose $\beta \in [0.8,1]$ uniformly at random at every step. 
This produces high quality solutions in the presence of large variance in constructed solutions \cite{Feo95greedyrandomized}. 



\vspace{-0.1in}
 \subsubsection{Local Search}
 
A solution generated by \cons~is a preliminary one and may not necessarily have the best robustness. Thus, it is almost always beneficial to apply a local refinement procedure to each constructed solution. A local search algorithm works in an iterative fashion by successively replacing the current solution with a better one in the neighborhood of the current solution. It terminates when no better solution can be found. We describe our local search phase in Algorithm \ref{alg:local}. 

As the \rls~asks for a subgraph of size $s$, the local search takes as input an $s$-subgraph generated by construction and searches for a better solution around it by ``swapping'' nodes in and out.
Ultimately it finds a locally optimal subgraph of size upper bounded by ($s$+$1$).
As an answer, it returns the best $s$-subgraph with the highest robustness found over the iterations.
As such, \local~is an adaptation of a general local search procedure to yield subgraphs of desired size.

%
%

\vspace{-0.1in}
\begin{algorithm}[h]
\caption{\local} \label{alg:local}
\begin{algorithmic}[1]
\REQUIRE Graph $G = (V,E)$, $S$, integer $s$
\ENSURE Subset of nodes $S' \subseteq V$, $|S'|=s$
\STATE $more \leftarrow$ TRUE, ${S'} \leftarrow S$
\WHILE{$more$} 
	\IF {$\exists v\in S$ such that
		$\bar{\lambda}(S\backslash \{v\}) \geq \bar{\lambda}(S)$}  
		\STATE $S := S \backslash \{v^*\}$ s.t. 
		$v^*:= \max_{v\in {\cal N}(S)\backslash S}  \bar{\lambda}(S\backslash \{v\})$
		\LINEIFONE{$|S|=s$}{${S'} \leftarrow S$}
	\ELSE
	\STATE  $more \leftarrow$ FALSE
	\ENDIF
	
	\STATE $add \leftarrow$ TRUE	
	\WHILE{$add$ and $|S|\leq s$}
	\IF {$\exists v\in {\cal N}(S)\backslash S$  s.t.
		$\bar{\lambda}(S\cup \{v\}) > \bar{\lambda}(S)$}  
		\STATE $S := S \cup \{v^*\}$,
		$v^*:= \max_{v\in {\cal N}(S)\backslash S}  \bar{\lambda}(S\cup \{v\})$  
		\STATE  $more \leftarrow$ TRUE
		\LINEIFONE{$|S|=s$}{${S'} \leftarrow S$}
		\ELSE
		\STATE {$add \leftarrow$ FALSE}
	\ENDIF
	\ENDWHILE
	
\ENDWHILE
\RETURN ${S'}$
\end{algorithmic}
\end{algorithm}
\vspace{-0.1in}

The local search is guaranteed to terminate, as the objective value (i.e., subgraph robustness) improves with every iteration and it is upper-bounded 
by the robustness of the $(s+1)$-clique. 
We provide the complexity analysis and the \gmrs~algorithm variants in Appendix \ref{sec:graspp}.

\vspace{-0.05in}
\section{Evaluation}
\label{sec:experiments}

We evaluate our methods extensively on numerous synthetic and real-world graphs.
Our real graphs, as in Table \ref{tab:datum}, come from various domains, including biological, email, Internet AS backbone, P2P, collaboration, and the Web.

\begin{table}[!t]
\vspace{-0.2in}
\caption{\small Real-world graphs. $\delta$: edge density, $\bar{\lambda}$: robustness}
{\small {
\vspace{-0.2in}
\label{tab:data}
\begin{center}
\begin{tabular}{lrrrr}
\hline
\textbf{Dataset} & \textbf{$n=|V|$} & \textbf{$m=|E|$} & \textbf{$\delta$} & \textbf{$\bar{\lambda}$} \\
\hline\hline
\textit{Jazz}	 		& 198  & 2742  &0.1406 &34.74 \\
\textit{Celegans N.}		& 297  & 2148  &0.0489 &21.32 \\
\textit{Email}			& 1133 & 5451 &0.0085 &13.74 \\ \hline
{\em Oregon-A} & 7352 & 15665 & 0.0005 & 42.29\\ 
{\em Oregon-B} & 10860 & 23409 & 0.0004&  47.54\\ 
{\em Oregon-C} & 13947 & 30584 & 0.0003 & 52.10\\ \hline
{\em P2P-GnutellaA}& 6301 & 20777 & 0.0010 & 19.62\\ 
{\em P2P-GnutellaB}& 8114 & 26013 & 0.0008 & 19.45\\ 
{\em P2P-GnutellaC}& 8717 & 31525 & 0.0008 & 13.35  \\ 
{\em P2P-GnutellaD}& 8846 & 31839 & 0.0008 & 14.46\\ 
{\em P2P-GnutellaE}& 10876 & 39994 & 0.0007 & 7.83\\ \hline
\textit{DBLP} & 317080 & 1049866 &2.09$\times 10^{-5}$& 103.18 \\
\textit{Web-Google}      & 875713 & 4322051& 1.13$\times 10^{-5}$& 99.36 \\
\hline
\end{tabular}
\vspace{-0.3in}
\end{center}
}}
\label{tab:datum}
\end{table}

Our work is in the general lines of subgraph mining, however with a new objective based on robustness. 
The closest to our setting is the densest subgraph mining. Therefore, we compare our results to dense subgraphs found by Charikar's algorithm \cite{conf/approx/Charikar00} (which we refer to as Charikar), as well as by Tsourakakis {\em et al.}'s two algorithms \cite{conf/kdd/TsourakakisBGGT13} (which we refer to as Greedy and LS for local search following the convention in their work).
We remark that the objectives used in those works are distinct; namely, average degree and edge-surplus, respectively, and are also different from ours.


We first evaluate the accuracy of the algorithms against ground truth. To do so, we create synthetic graphs and inject a clique in each graph. Note that a clique is both the densest and the most robust subgraph of a certain size. Therefore, the algorithms are compared on the same grounds. 

Table \ref{tab:er} provides precision, recall, and subgraph size $|S|$ averaged over ten Erd\H{o}s-R\'{e}nyi random graphs, with $n=3000$ nodes and $p=\{0.5,0.1,0.008\}$, in which we inject a clique of size $30$.  $p$'s are selected to capture very dense, medium-dense, and sparse graphs.
We notice that while all methods perform sufficiently well for sparse graphs with $p=0.008$, accuracy of \gmrs~is superior to the competing methods at all densities.

\begin{table*}
\caption{\small $P$recision\& $R$ecall (avg.) for our \gmrs \& \greedymrs, Charikar \protect\cite{conf/approx/Charikar00}, Greedy \& LS \protect\cite{conf/kdd/TsourakakisBGGT13} on ten ER graphs.
}
\vspace{-0.1in}
{\small {
\begin{center}
\begin{tabular}{ll|rr|rr|rrr|rrr|rrr}
\hline
\multicolumn{2}{c|}{ER parameters} 
& \multicolumn{2}{c|}{\gmrs} 
& \multicolumn{2}{c|}{\greedymrs} 
& \multicolumn{3}{c|}{Charikar \protect\cite{conf/approx/Charikar00}} 
& \multicolumn{3}{c|}{Greedy \protect\cite{conf/kdd/TsourakakisBGGT13}} 
& \multicolumn{3}{c}{Local Search \protect\cite{conf/kdd/TsourakakisBGGT13}} 
\\
 \hline
$n$ & $p$ & $|S|$ & $P=R$ & $|S|$ & $P=R$ & $|S|$ & $P$ & $R$ & $|S|$ & $P$ & $R$ & $|S|$ & $P$ & $R$ \\
 \hline
3000 & 0.5&	30&	0.97&	30	& 0.02	&	3000 & 0.01&	1 &	3000 & 0.01	& 1	& 3000 &	0.01&	1\\
3000 & 0.1 & 30 & 1 & 30 & 0.95 & 3000 & 0.01 & 1 & 29.60& 0.99 &  0.97 & 20.63 &  0.37 & 0.35 \\
3000 & 0.008 & 30 & 1 & 30 & 0.99 & 30 & 1 & 1 & 30 & 1 & 1 & 28.23 & 0.94 & 0.93 \\
\hline    
\end{tabular}
\end{center}
}}
\label{tab:er}
\vspace{-0.25in}
\end{table*}

\begin{figure}[t!]
\vspace{-0.05in}
\centering
\begin{tabular}{cc}
\hspace{-0.1in}\includegraphics[width=1.65in,height=1.35in]{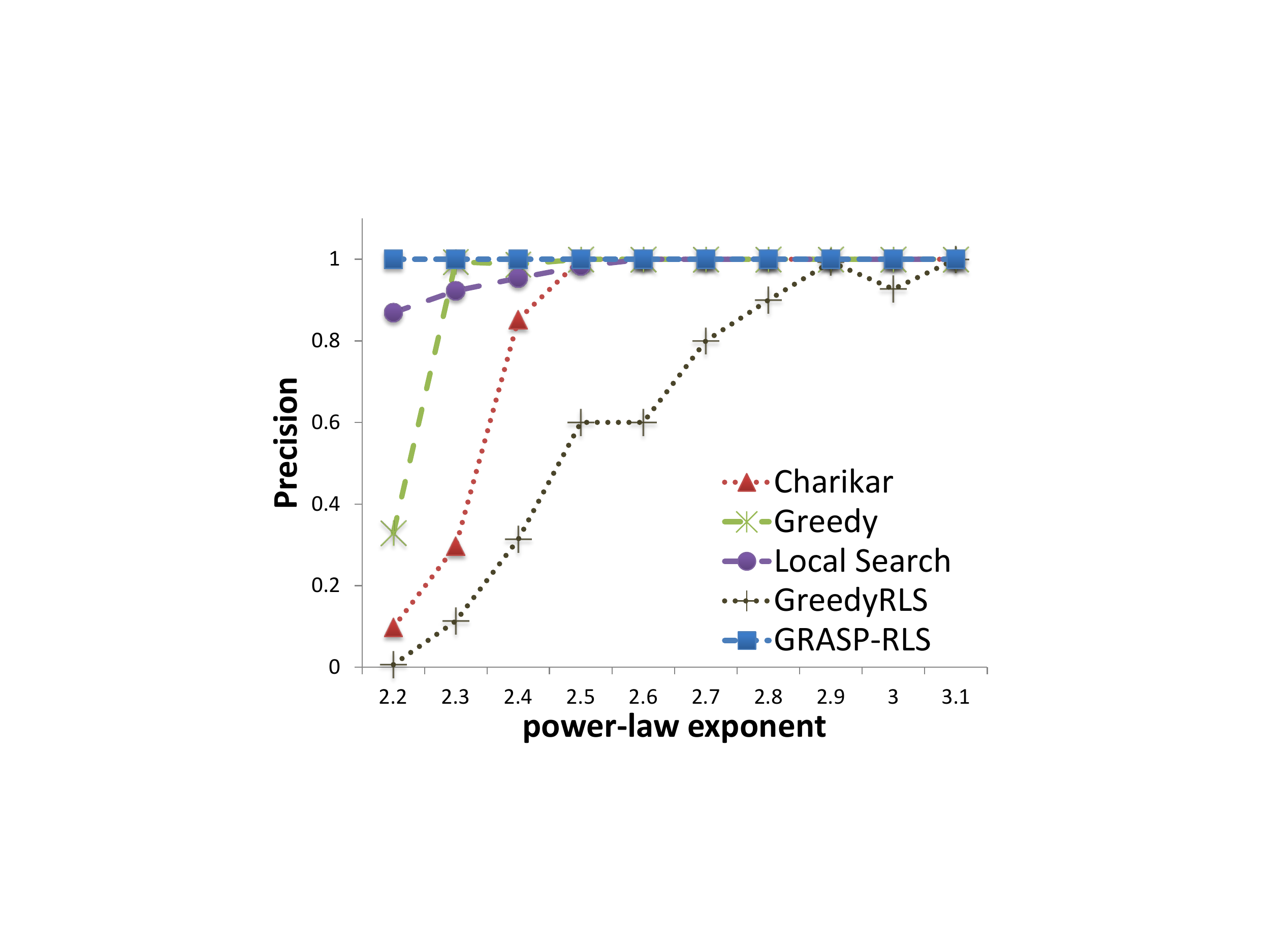} &
\hspace{-0.1in}\includegraphics[width=1.65in,height=1.35in]{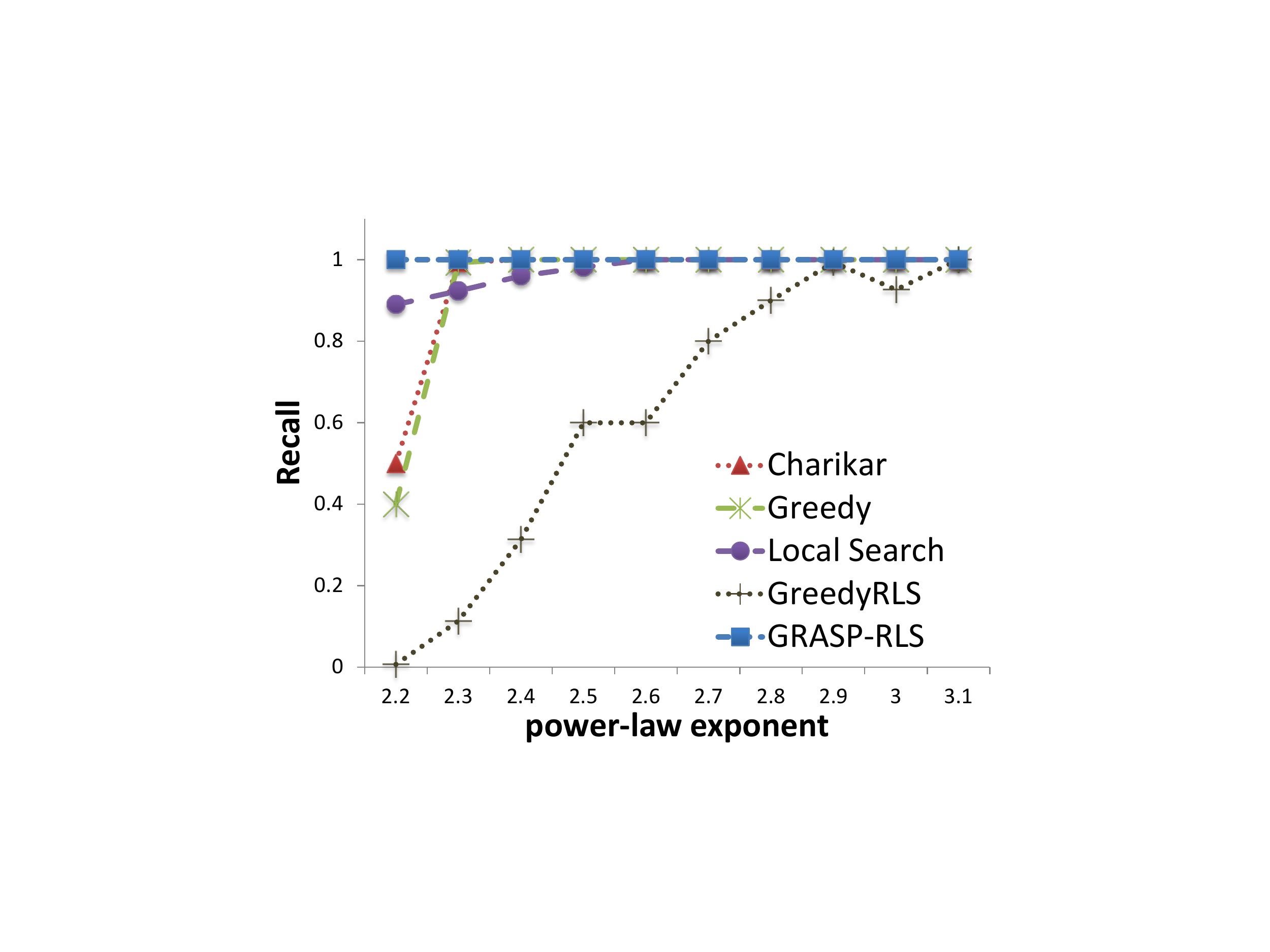} \\
\end{tabular}
\vspace{-0.15in}
\caption{\small Precision \& Recall for our \gmrs, Charikar \protect\cite{conf/approx/Charikar00}, Greedy \& Local Search \protect\cite{conf/kdd/TsourakakisBGGT13} vs. exponent of power-law graphs.} 
\label{fig:precall}
\vspace{-0.15in}
\end{figure}

We also compare the algorithms on the Chung-Lu random power-law graphs \cite{ChungLu03}, with $n=3000$ and power-law exponent varying from $2.2$ to $3.1$ as observed in real graphs (larger exponent implies a sparser graph), in which we inject a clique of $15$ nodes. 
We run the LS algorithm seeded with one
of the nodes of the clique as previously done in \cite{conf/kdd/TsourakakisBGGT13}, while \gmrs~is not favored by such a selection.
  Precision and recall curves averaged over ten graphs are given in Figure \ref{fig:precall}. We note that while the accuracies of all methods improve by the increasing exponent as the task becomes easier, \gmrs~remains superior with robust performance at all host graph densities.

Cliques are both the densest and the most robust subgraphs, however, it is expected that the algorithms will find different subgraphs in general due to their distinct objectives. To understand their differences, we turn to real world graphs
and compare the robust and dense subgraphs based on three main criteria: (a) robustness $\bar{\lambda}$ as in Equ. \eqref{natur}, (b) triangle density ${t[S]}/{\binom{|S|}{3}}$, and (c) edge density ${e[S]}/{\binom{|S|}{2}}$.

Table \ref{tab:results} shows results on our largest graphs from each category.
Note that the three algorithms we compare to try to find the densest subgraph without a size restriction. Thus, each one obtains a subgraph of a different size.
To make the robust subgraphs (RS) comparable to the densest subgraphs (DS), we find subgraphs of size $s$ equal to the ones found by Charikar, Greedy, and LS, respectively noted as $s_{Ch}$, $s_{Gr}$, and $s_{Ls}$.
As such, we compare to the \textit{best} results achieved by each of the densest subgraph algorithms.



\begin{table*}[t!]
\caption{\small Comparison of robust and densest subgraphs. Ch: Charikar \protect\cite{conf/approx/Charikar00}, Gr: Greedy \protect\cite{conf/kdd/TsourakakisBGGT13}, Ls: Local search \protect\cite{conf/kdd/TsourakakisBGGT13}.
}
\vspace{-0.1in}
{\small {
\begin{center}
\begin{tabular}{c|l||rrr|rrr|rrr}
\hline
Data
& 
Method
& \multicolumn{3}{c|}{robustness $\bar{\lambda[S]}$} 
& \multicolumn{3}{c|}{triangle density $\Delta[S]$} 
& \multicolumn{3}{c}{edge density $\delta[S]$} 
\\
 \hline\hline
& ($s_{Ch}$, $s_{Gr}$, $s_{Ls}$) &  Ch    &   Gr    &   Ls    &   Ch    &   Gr    &  Ls    &   Ch    &    Gr   &   Ls     \\ 
 \hline
\parbox[t]{2mm}{\multirow{3}{*}{\rotatebox[origin=c]{90}{\textit{Email}}}}
& DS (271, 12, 13) &   13.58    &   8.51   &  4.96   &   \textbf{0.0009}    &  \textbf{1.0000}     &   0.2237   &    \textbf{0.0600}   &  \textbf{1.0000}     &   0.5897     \\
& RS-{\sc Greedy} &  13.94     &    5.96   & 6.27    &   0.0001    &  0.0696     &  0.0606    &   0.0523    &   0.7576    &   0.7179  \\
& RS-{\sc GRASP} &   \textbf{14.04}    &   \textbf{8.52}    &  \textbf{8.91}    &   0.0007    &  \textbf{1.0000}     &  \textbf{0.8671}    &  0.0508     &       \textbf{1.0000}&  \textbf{0.9487}   \\
\hline
\parbox[t]{2mm}{\multirow{3}{*}{\rotatebox[origin=c]{90}{{\em Oreg-C}}}} 
& DS (87, 61, 52) &   34.44    &    30.01   &   27.69   &   0.0868    &   0.1768    &  0.2327    &  \textbf{0.3892}     &  \textbf{0.5311}     &  0.5927     \\
& RS-{\sc Greedy} &   34.31    &    24.70   &   21.75   &   0.0857    &   0.1022    &  0.1193    &  0.3855     &       0.4131		&  0.4367   \\
& RS-{\sc GRASP} &   \textbf{34.47}    &  \textbf{30.14}     &  \textbf{28.01}    &   \textbf{0.0870}    &   \textbf{0.1775}    &   \textbf{0.2375}   &  0.3884     &  0.5301     &  \textbf{0.5943}    \\
\hline    
\parbox[t]{2mm}{\multirow{3}{*}{\rotatebox[origin=c]{90}{{\em P2P-E}}}} 
& DS (386, 22, 4)&    8.81   &    6.40   &   0.86   &  \textbf{9.77E-06}     &  0.0    &   0.0   &  \textbf{0.0306}     &  \textbf{0.4372}     &  0.6667     \\
& RS-{\sc Greedy} &   9.10    &    5.22   &  0.86    &   6.83E-06    &   0.0    &  0.0    &   0.0267   &  0.3593     & 0.6667    \\
& RS-{\sc GRASP} &   \textbf{9.22}    &   \textbf{6.41}    &   \textbf{1.29}   &   6.93E-06     &  0.0    &   \textbf{0.5}   &  0.0270     &  \textbf{0.4372}     & \textbf{0.8333}     \\
\hline    
\parbox[t]{2mm}{\multirow{3}{*}{\rotatebox[origin=c]{90}{{\em Web}}}} 
& DS (240, 105, 18) &   52.15    &  47.62     &   10.20   &   \textbf{0.0266}    &   \textbf{0.2160}    &   0.4178   &  \textbf{0.2274}     &  \textbf{0.4759}     &  0.7254     \\
& RS-{\sc Greedy} &   41.57    &   22.56    &  8.69    &   0.0027    &    0.0082   &   0.2525   &  0.0710     &  0.1225     &  0.6144   \\
& RS-{\sc GRASP} &    \textbf{53.96}   &   \textbf{48.68}    &  \textbf{14.11}    &  0.0153     &  0.1246     &   \textbf{1.0000}   &  0.1296     &  0.3996     &  \textbf{1.0000}    \\
\hline    
\end{tabular}
\end{center}
}}
\vspace{-0.25in}
\label{tab:results}
\end{table*}

We notice that densest subgraphs found by Greedy and LS are often substantially smaller than those found by Charikar, and also have higher edge density, which is the same observation as in \cite{conf/kdd/TsourakakisBGGT13}.
On the other hand, robust subgraphs have higher robustness than densest subgraphs, {\em even} at lower densities.
This shows that high density does not always imply high robustness, and vice versa, illustrating the differences in the two problem settings. 

Thus far, we also note that \gmrs~consistently outperforms \greedymrs, suggesting that the proposed bottom-up search is superior to the greedy top-down search.

\begin{figure}[h!]
\vspace{-0.1in}
\centering
\includegraphics[width=0.4\textwidth, height=1.95in]{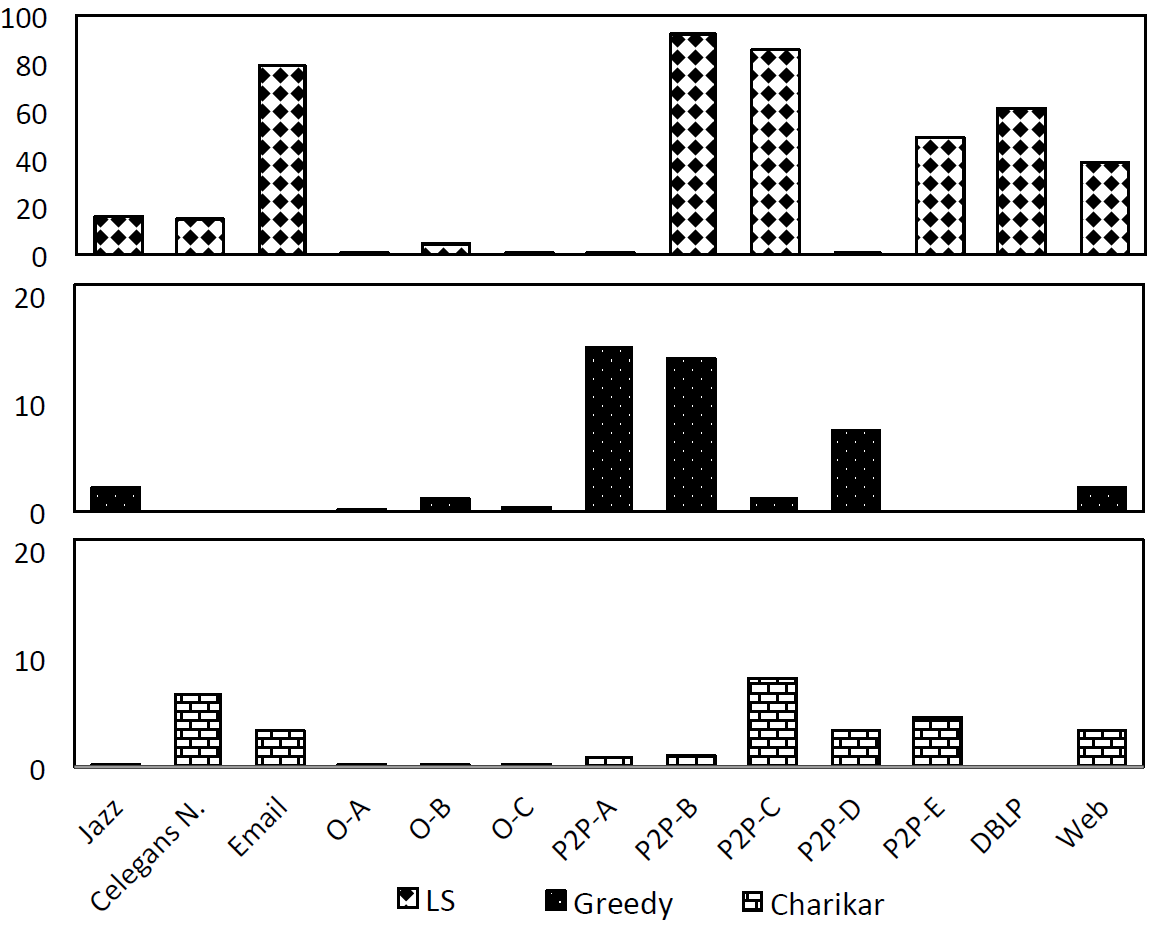}
\vspace{-0.15in}
\caption{\small Robustness improvement (\%) of \gmrs~over (top to bottom) best LS, Greedy, and Charikar results.} 
\label{fig:percent}
\vspace{-0.05in}
\end{figure}

Figure \ref{fig:percent} shows the relative difference in robustness of \gmrs~subgraphs over again, the best results obtained by Charikar, Greedy, and LS on all of our real graphs. We achieve a wide range of improvements depending on the graph, where the difference is always positive. The improvements with respect to the LS results are the most pronounced.

\begin{figure}[!t]
\vspace{-0.05in}
\centering
\begin{tabular}{cc}
\hspace{-0.1in}\includegraphics[width=1.62in,height=1.27in]{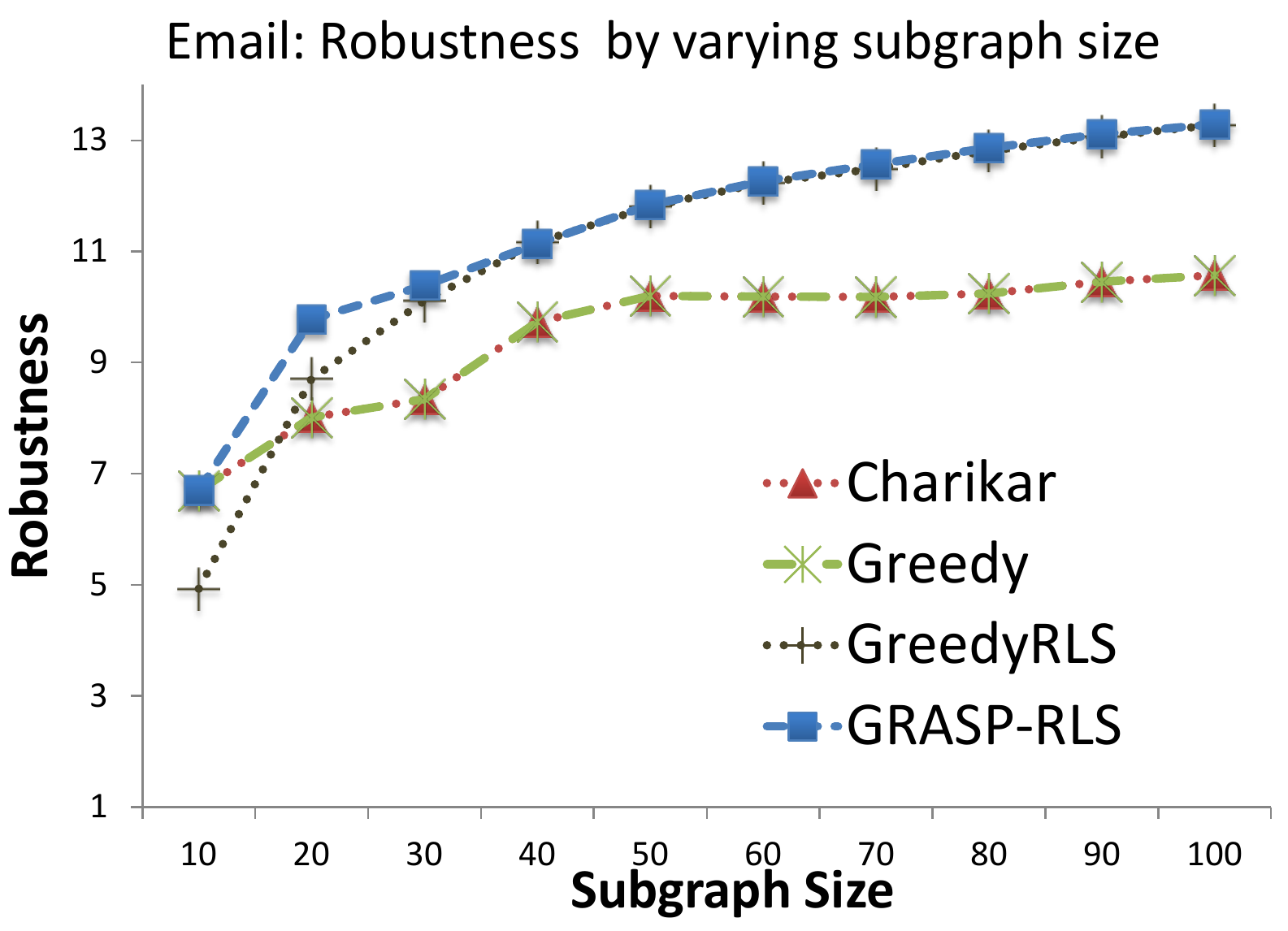} &
\hspace{-0.1in}\includegraphics[width=1.62in,height=1.27in]{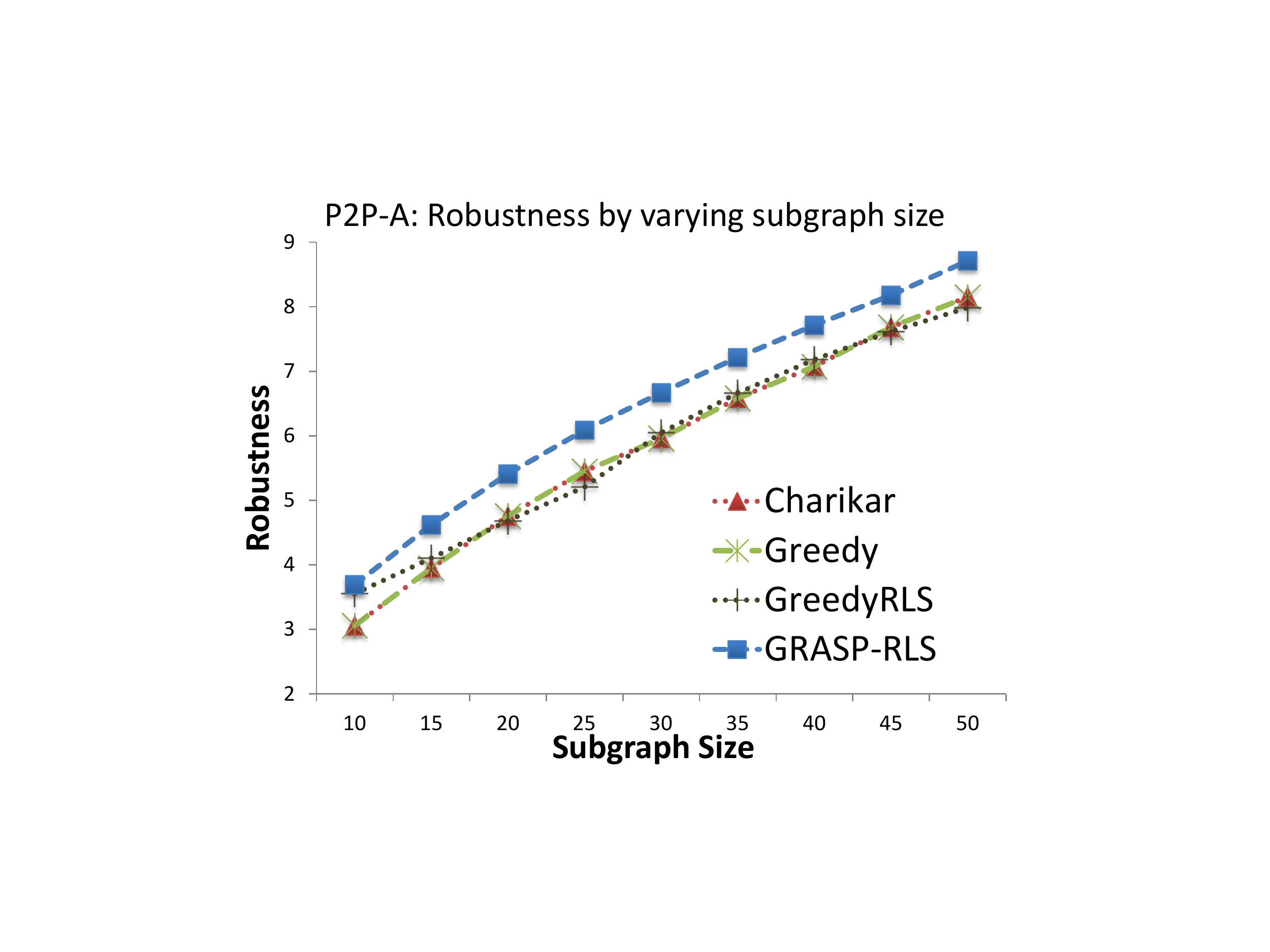} \\
\end{tabular}
\vspace{-0.2in}
\caption{\small Subgraph robustness at varying subgraph sizes $s$.} 
\label{fig:robvss}
\vspace{-0.2in}
\end{figure}

Comparisons in Table \ref{tab:results} and Figure \ref{fig:percent} are for subgraphs at sizes where best results are obtained for each of the three densest subgraph algorithms. Our algorithms, on the other hand, accept a subgraph size input $s$. Thus, we next compare the competing methods at varying output sizes. Charikar and Greedy are both top-down methods, in which the lowest degree node is removed at each step and the best subgraph (best average degree or edge surplus, respectively) is output among all graphs created along the way. We modify these so that we pull out the subgraphs when size $s$ is reached during the course of their run.\footnote{\scriptsize Local search by \cite{conf/kdd/TsourakakisBGGT13} finds locally optimal subgraphs, which are not guaranteed to grow to a given size $s$. Thus, we omit comparison to LS subgraphs at varying sizes. Figure \ref{fig:percent} shows that improvements over LS subgraphs are already substantially large.} 
Figure \ref{fig:robvss} shows that our \gmrs~produces  subgraphs with higher robustness at varying sizes on two example graphs (similar results on others). This also shows that the densest subgraph approaches are not directly applicable to our problem.

%
%


Experiments thus far illustrate that we find subgraphs with robustness higher than the densest subgraphs. 
These are relative results. To show that the subgraphs we find are in fact robust, we next
quantify the magnitude of the robustness values we achieve through significance tests. 

Given a subgraph that \gmrs~finds, we bootstrap $B=1000$ new subgraphs by rewiring its edges at random. We compute an empirical $p$-value for each subgraph by dividing the number of randomly rewired subgraphs that have larger robustness by $B$.
The $p$-value essentially captures the probability that we would be able to obtain a subgraph with robustness greater than what we find by chance if we were to create a topology with the same number of nodes and edges at random (note that all such subgraphs would have the same edge density). Thus a low $p$-value implies that, among the same density topologies, the one we find is in fact robust with high probability. 

\begin{figure}[h]
\vspace{-0.1in}
\centering
\includegraphics[width=0.45\textwidth]{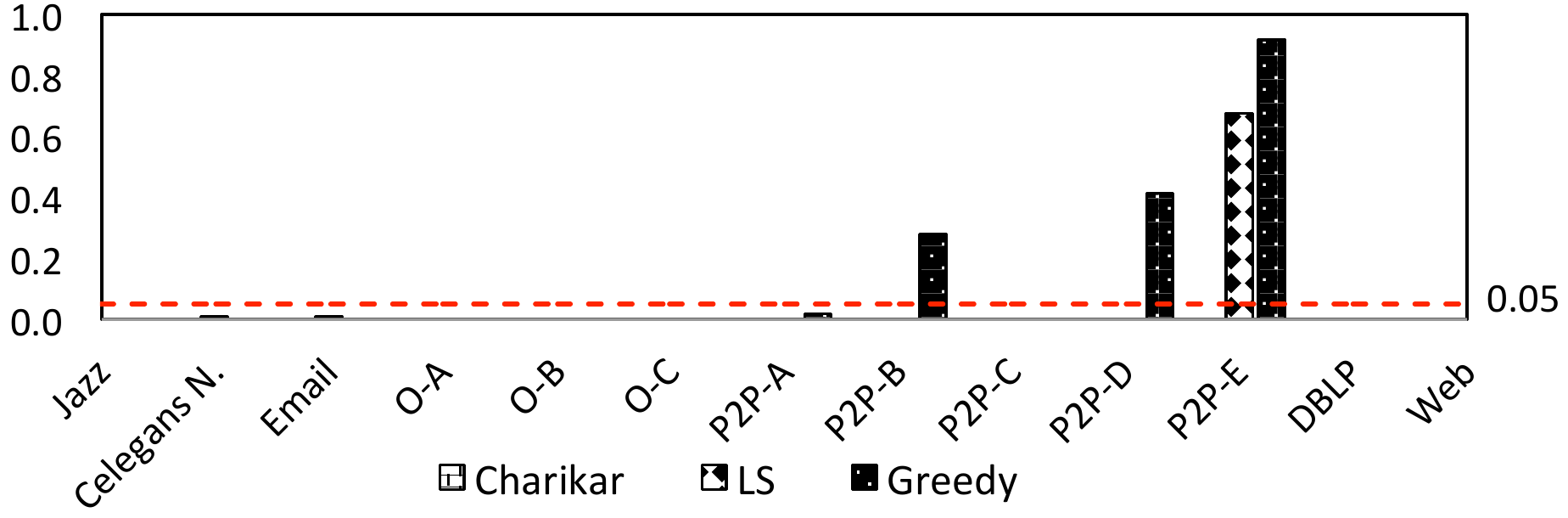}
\vspace{-0.15in}
\caption{\small $p$-values of significance tests indicate that \gmrs~subgraphs have  significantly large robustness.} 
\label{fig:pval}
\vspace{-0.1in}
\end{figure}

Figure \ref{fig:pval} shows that the subgraphs we find on almost all real graphs are significantly robust at $0.05$. 
For cases with large $p$-values, it is possible to obtain higher robustness subgraphs with rewiring. For example, {\em P2P-E} is a graph where all the robust subgraphs (also the dense subgraphs) found contain very few or no triangles (see Table \ref{tab:results}). Therefore, rewiring edges that short-cut longer cycles they contain help improve their robustness. We remark that large $p$-values indicate that the found subgraphs are not significantly robust, but does not imply our algorithms are unable to find robust subgraphs. That is because the rewired more robust subgraphs do not necessarily exist in the original graph $G$, and it is likely that $G$ does not contain any subgraph with robustness that is statistically significant.

%
%
%

%

%
%

\begin{figure}[h!]
\vspace{-0.1in}
\centering
\begin{tabular}{cc}
\hspace{-0.1in}\includegraphics[width=1.65in,height=1.265in]{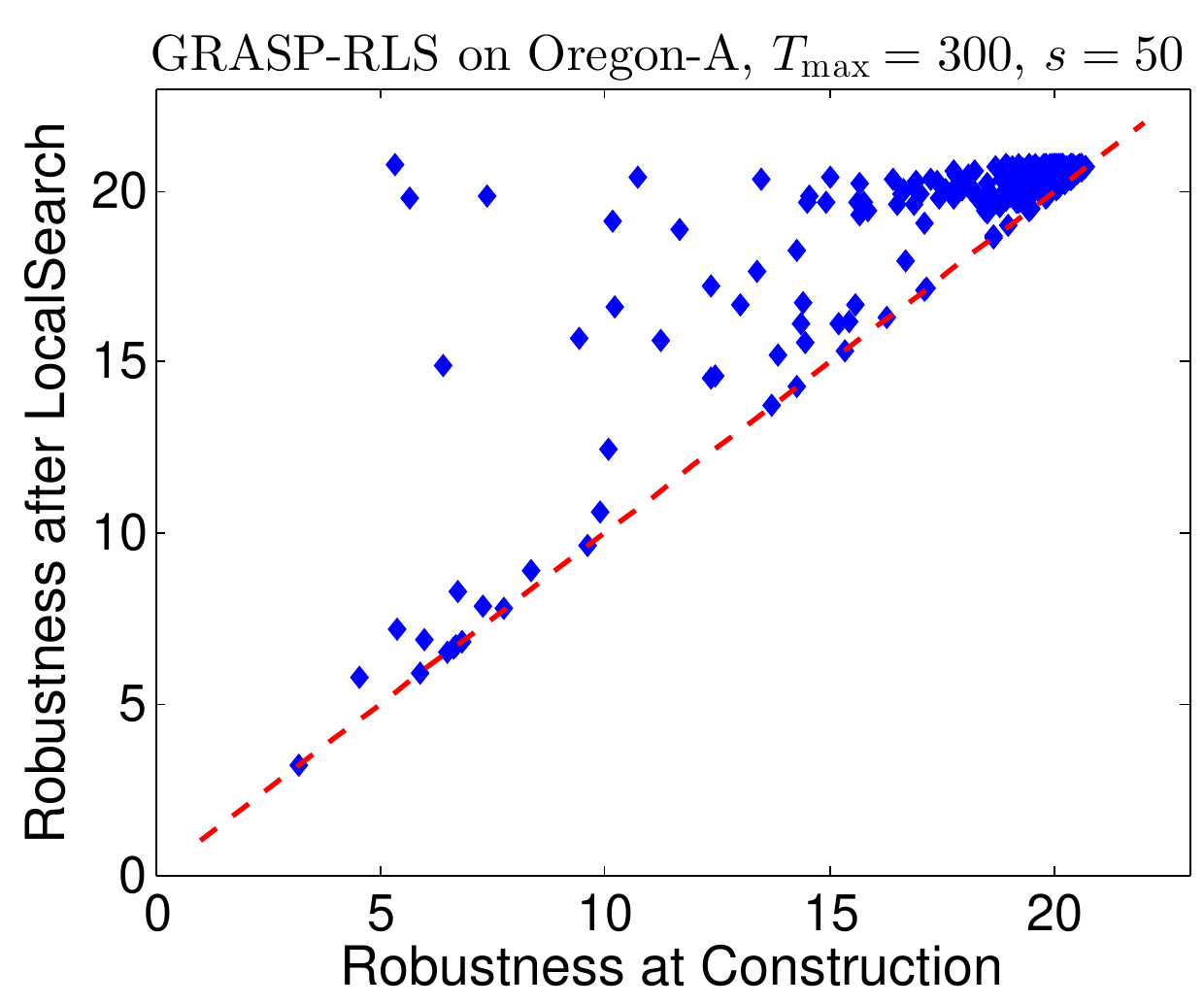} &
\hspace{-0.1in}\includegraphics[width=1.65in,height=1.25in]{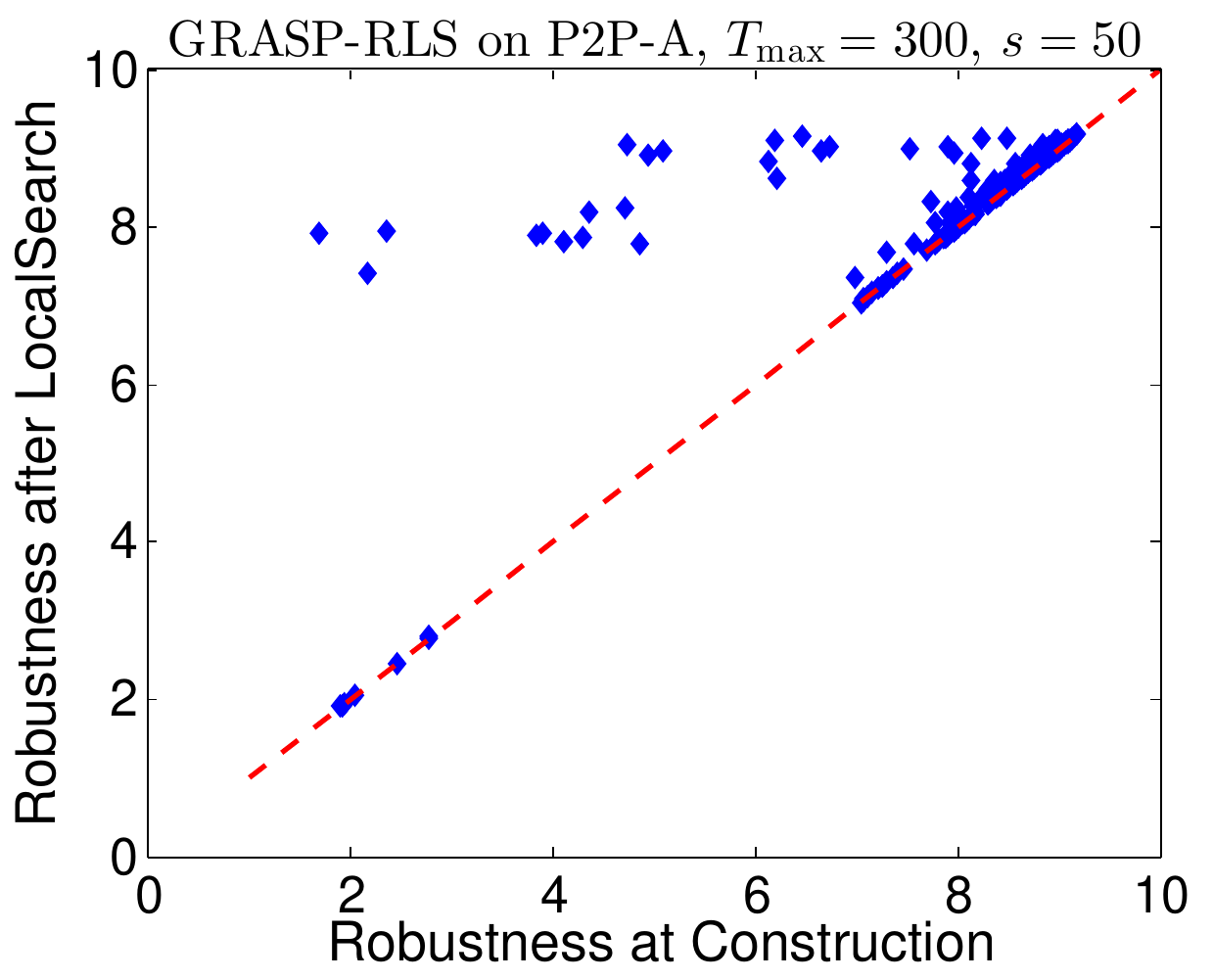} \\
\end{tabular}
\vspace{-0.2in}
\caption{\small $\bar{\lambda}$ achieved at \cons~versus after \local.} 
\label{fig:improve}
\vspace{-0.05in}
\end{figure}

Next we analyze the performance of our \gmrs~approach in more detail. Recall that \cons~quickly builds a subgraph which \local~uses to improve over to obtain a better result. In Figure \ref{fig:improve} we show the robustness of subgraphs obtained at construction and after local search on two example graphs for $s=50$ and $T_{\max}=300$.
We notice that most of the \gmrs~iterations find a high robustness subgraph right at construction.
In most other cases, local search is able to improve over construction results significantly. 
In fact, the most robust outcome on {\em Oregon-A} (Figure \ref{fig:improve} left) is obtained when construction builds a subgraph with robustness around $\bar{\lambda}=6$, which the local search improves over $\bar{\lambda}=20$.

\begin{figure*}[!t]
\vspace{-0.25in}
\centering
\begin{tabular}{ccccc}
\hspace{-0.15in}\includegraphics[width=1.25in]{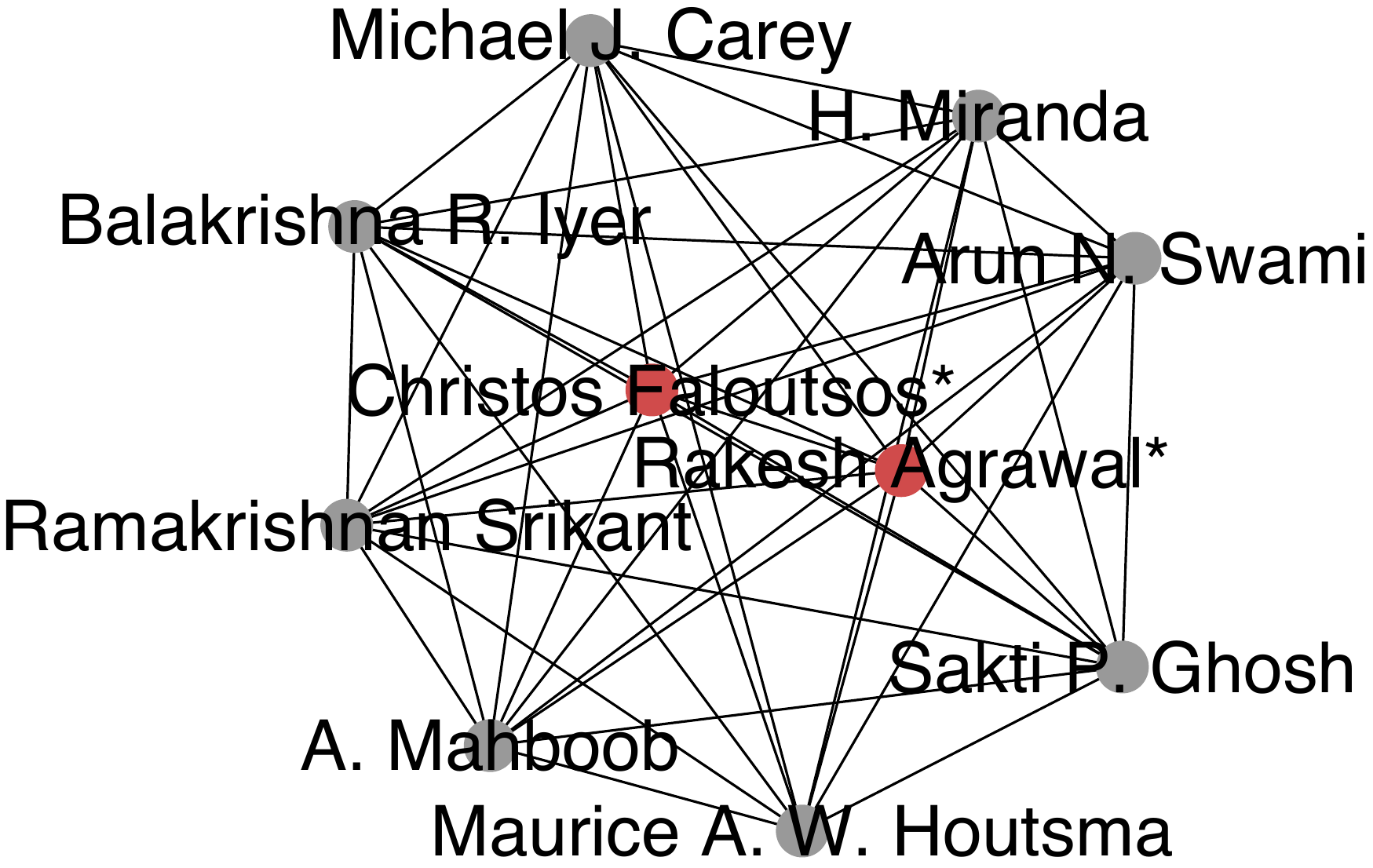} &
\hspace{-0.15in}\includegraphics[width=1.5in]{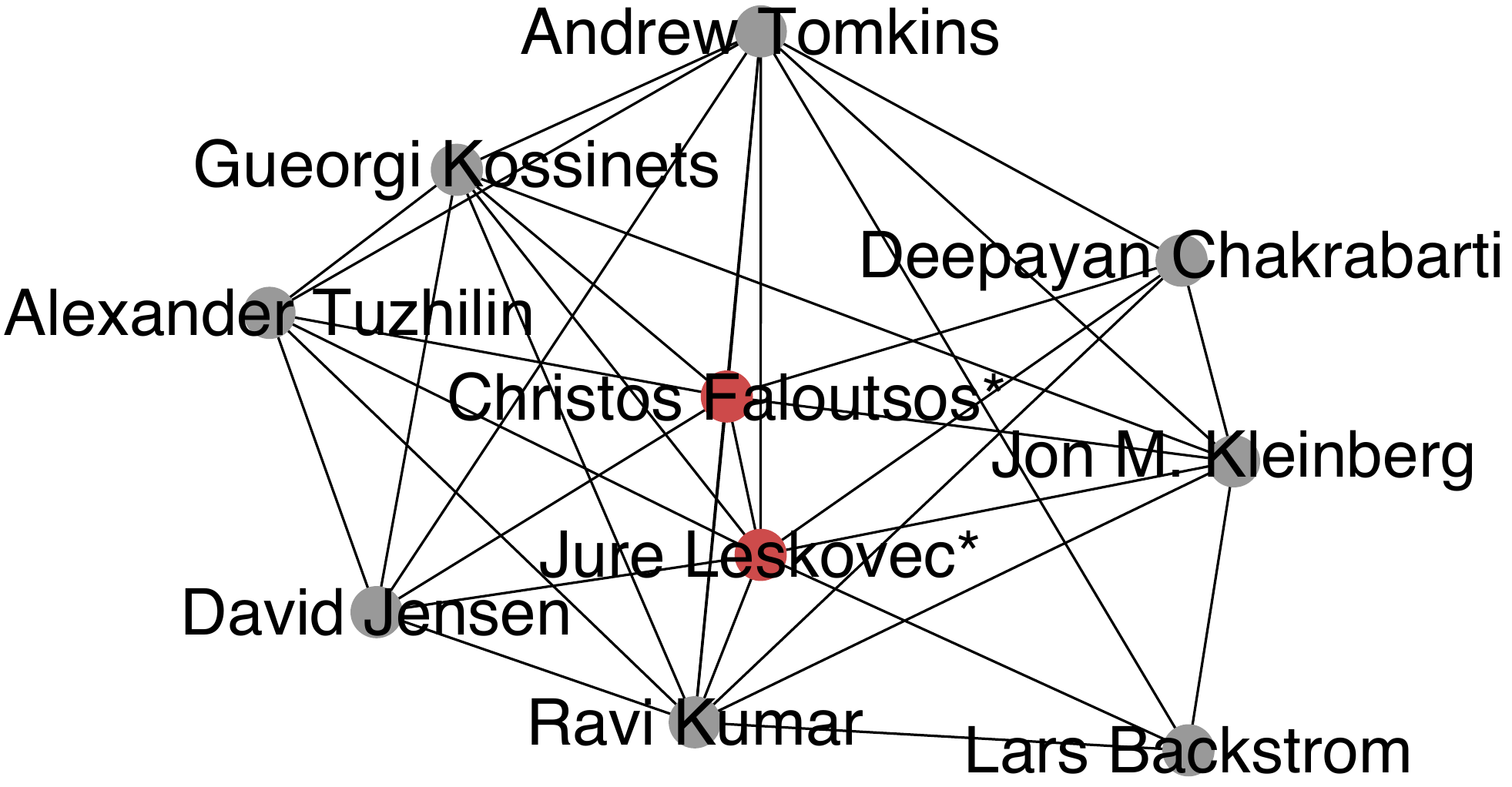} &
\hspace{-0.15in}\includegraphics[width=1.25in]{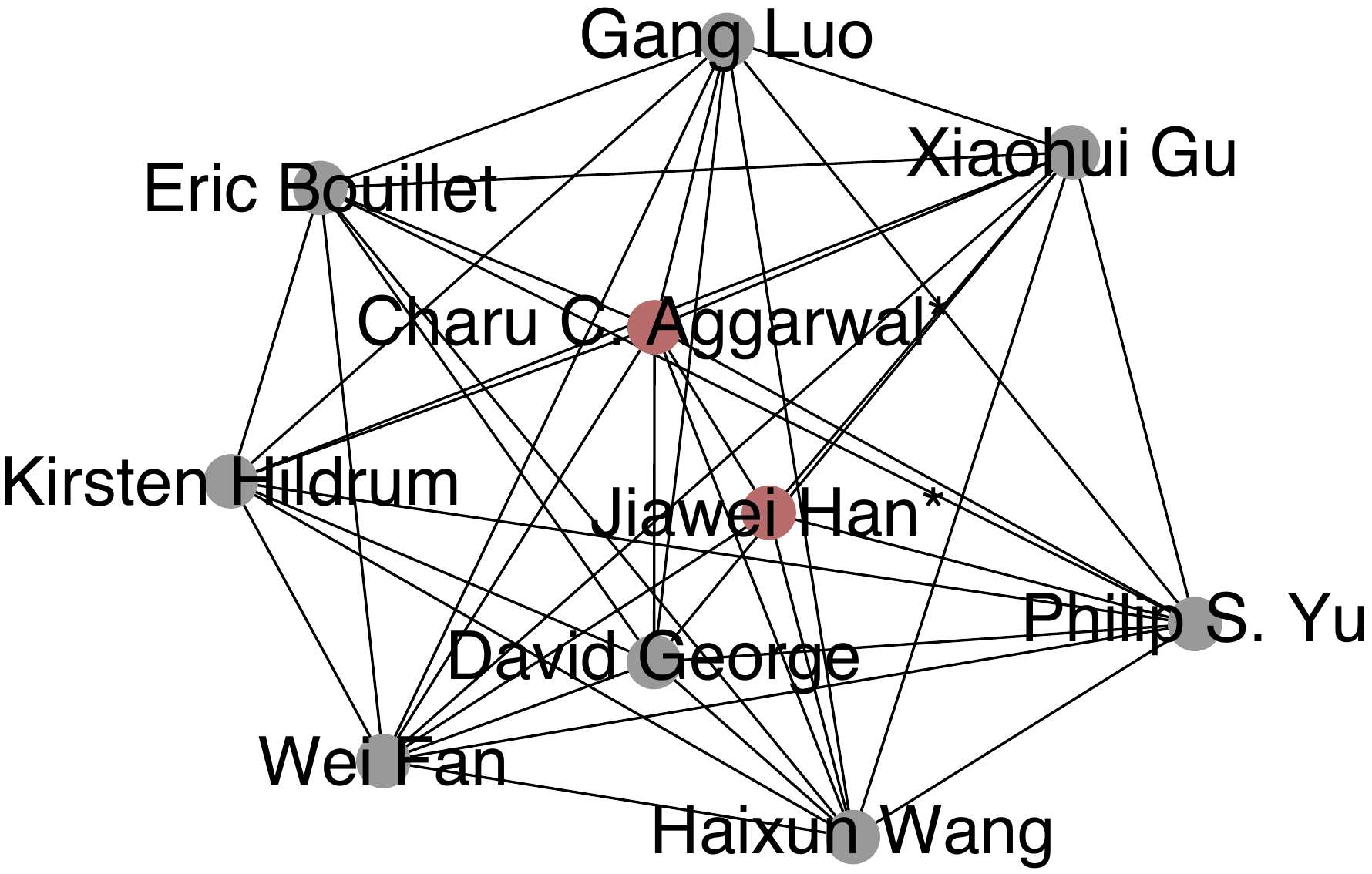} &
\hspace{-0.15in}\includegraphics[width=1.15in]{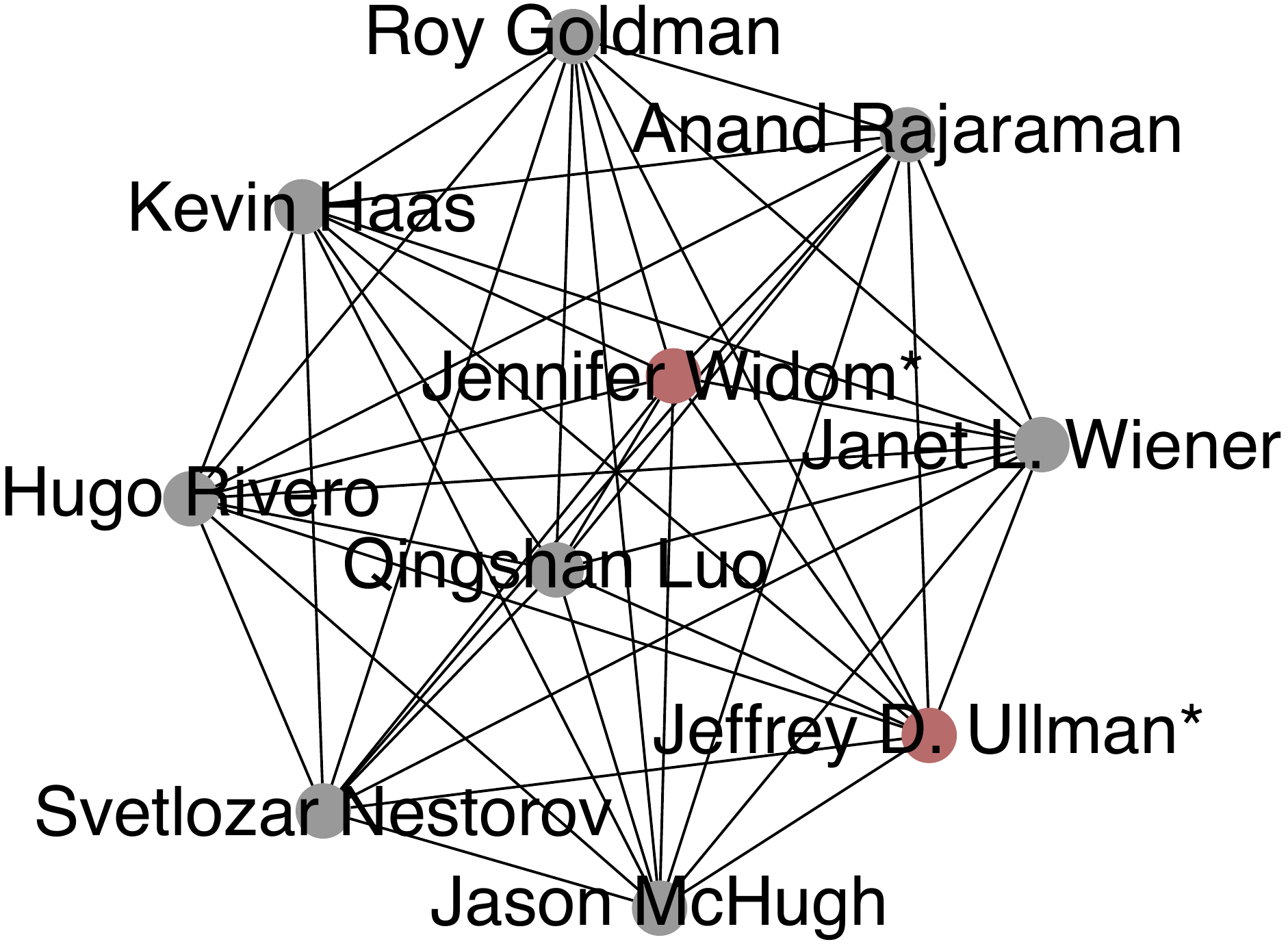} &
\hspace{-0.15in}\includegraphics[width=1.5in]{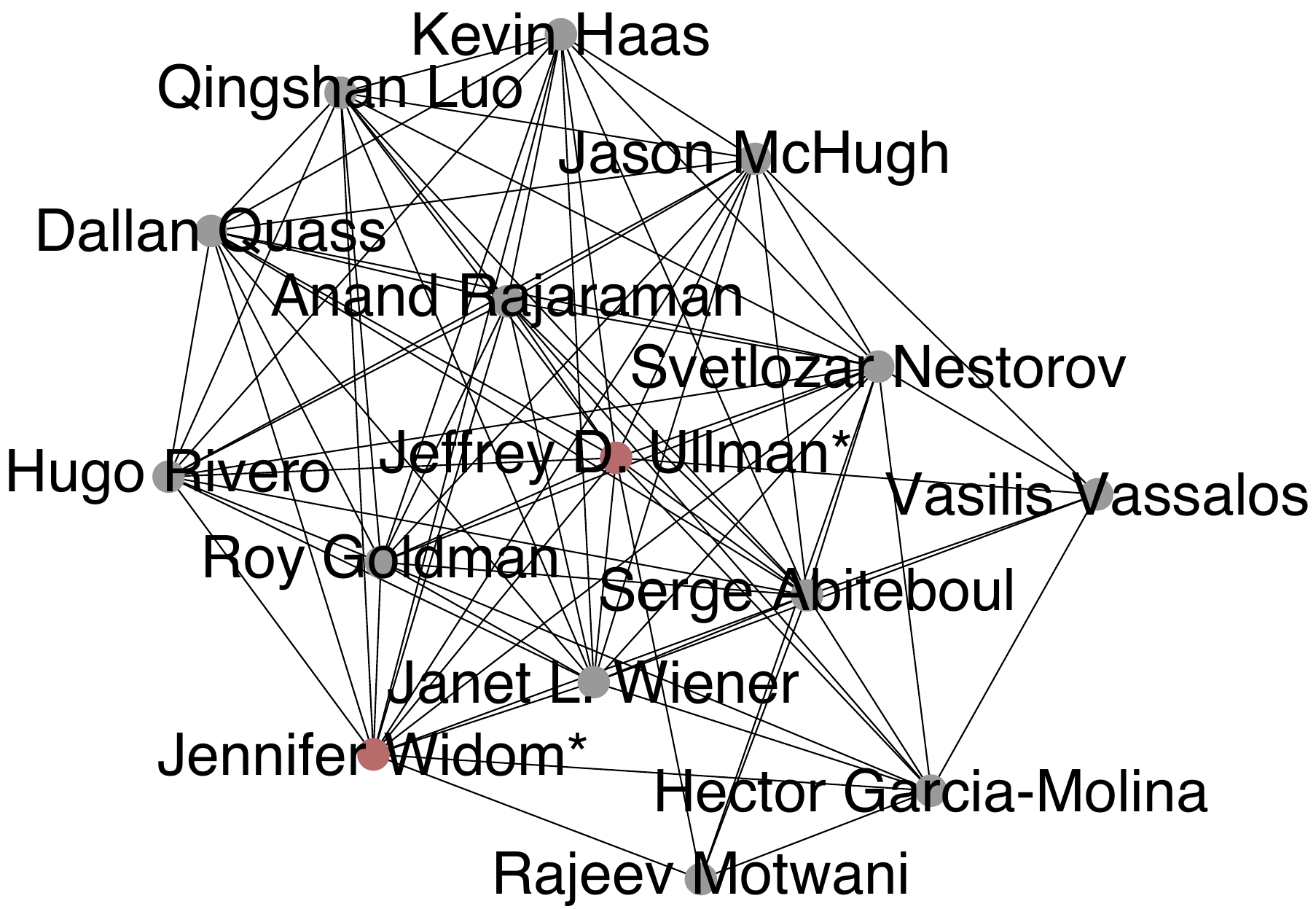} \\
\hspace{-0.15in} {\scriptsize  (a) \{C. Faloutsos, R. Agrawal\} }&
\hspace{-0.15in} {\scriptsize (b) \{C. Faloutsos, J. Leskovec\} }&
\hspace{-0.15in}  {\scriptsize (c) \{J. Han, C. C. Aggarwal\} }&
\multicolumn{2}{c}{\hspace{-0.15in}  {\scriptsize (d) \{J. Widom, J. D. Ullman\} }}\\
\hspace{-0.15in} {\scriptsize  $\delta$=$1$, $\Delta$=$1$, $\bar{\lambda}$=$6.7$} &
\hspace{-0.15in} {\scriptsize $\delta$=$0.78$, $\Delta$=$0.51$, $\bar{\lambda}$=$5.0$} &
\hspace{-0.15in} {\scriptsize $\delta$=$0.91$, $\Delta$=$0.78$, $\bar{\lambda}$=$6.0$} &
\hspace{-0.15in} {\scriptsize $\delta$=$1$, $\Delta$=$1$, $\bar{\lambda}$=$6.7$} &
\hspace{-0.15in} {\scriptsize $\delta$=$0.8$, $\Delta$=$0.58$, $\bar{\lambda}$=$9.0$} \\
\end{tabular}
\vspace{-0.1in}
\caption{\small Robust DBLP subgraphs returned by our \gmrs~algorithm when seeded with  authors indicated in (a)-(d). } 
\label{fig:viz}
\vspace{-0.15in}
\end{figure*}

We next perform several case studies on the DBLP co-authorship network to qualitatively analyze our subgraphs. Here, we use the seeded variant of our problem (Appendix \ref{sec:variants}).
Christos Faloutsos is a prolific researcher with various interests. 
In Figure \ref{fig:viz} (a), we invoke his interest in databases when used with Rakesh Agrawal as seeds, as Agrawal is an expert in this field.
Later in (b), we invoke his interest in data mining when we use Jure Leskovec as the second seed, who is a rising star in the field.
Likewise in (c) and (d) we find robust subgraphs around other selected prominent researchers in data mining and databases. In (d) we show how our subgraphs change with varying size. Specifically, we find a clique that the seeds J. Widom and J. Ullman belong to, for $s$=$10$. The subgraph of $s$=$15$, while no longer a clique, remains stable in which other researchers like R. Motwani and H. Garcia-Molina are included.

Given the local search characteristics of \gmrs, its complexity is linear in host graph size, as theoretically shown in Appendix \ref{sec:grasppcomp}.
Figure \ref{fig:scale} also illustrates the linear scalability w.r.t. input graph size empirically.\footnote{\scriptsize We use nine {\em Oregon} graphs with various sizes \cite{ChanSDM14}, the largest three of which are listed in Table \ref{tab:data}. Running time is averaged over $T_{\max}$ iterations as one can run each pair of construction followed by local search phases completely in parallel.}

\begin{figure}[h!]
\vspace{-0.1in}
\centering
\includegraphics[width=1.8in, height=0.35\textwidth, angle =270 ]{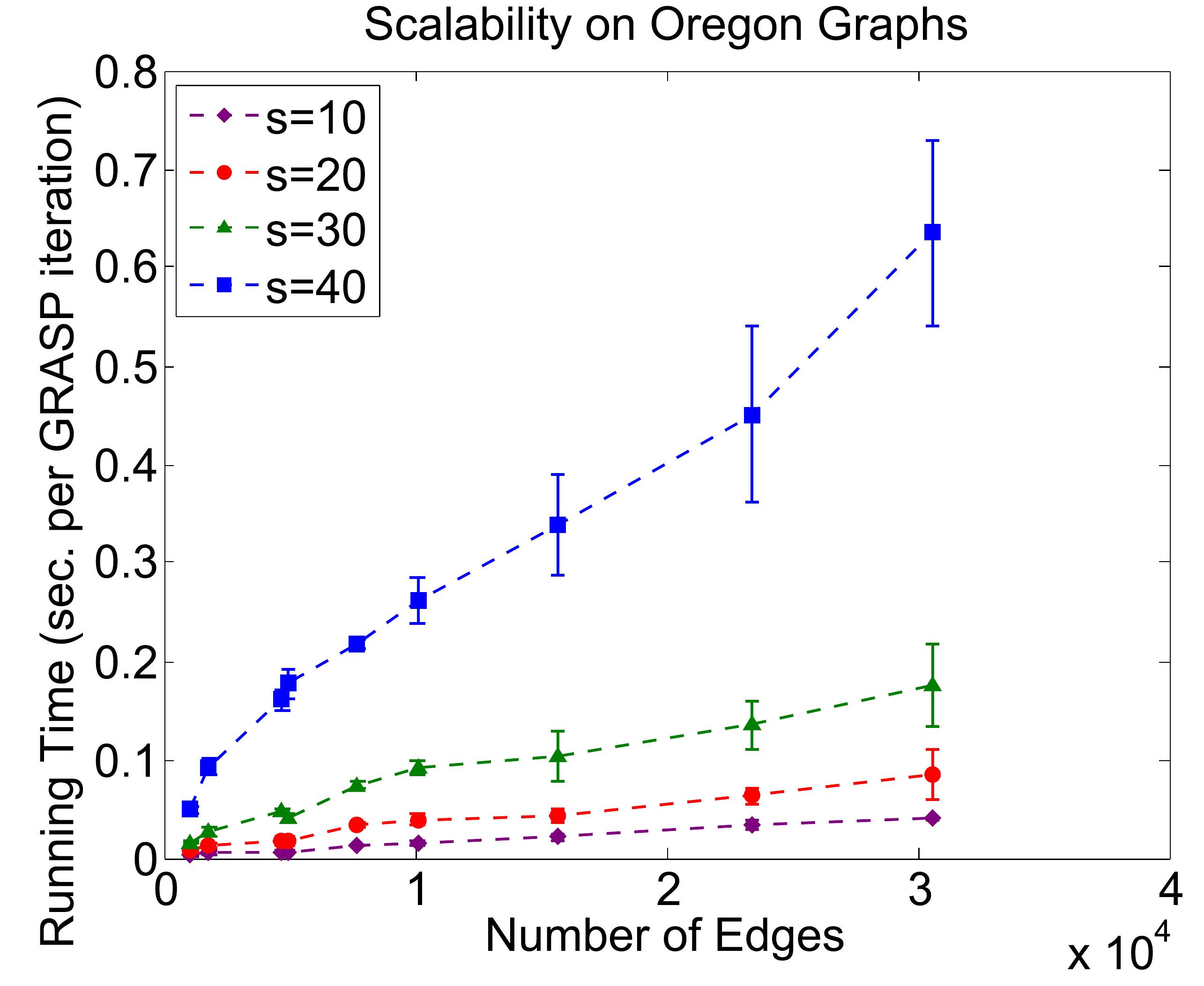}
\vspace{-0.15in}
\caption{\small Scalability of \gmrs~by graph size  $m$ and subgraph size $s$ (run time avg'ed over 10 runs, bars: 25\%-75\%).} 
\label{fig:scale}
\vspace{-0.15in}
\end{figure}

%
%
%
%
%
%
%
%


\section{Related Work}
\label{sec:related}

Robustness is a critical property of networked systems. Thus, it has been studied extensively in various fields including physics, biology, mathematics, and networking.
One of the early studies in measuring graph robustness shows that scale-free graphs are robust to random failures but vulnerable to intentional carefully-planned attacks \cite{albert2000error}.
This observation has stimulated studies on the response of networks to various attack strategies \cite{callaway2000nra,ChanSDM14,citeulike2352886,expansion2006,Holme2002,conf/cikm/TongPEFF12}.
Other works look at how to design networks that are optimal with respect to some survivability 
criteria \cite{FrankFrisch1970,journals/telsys/GhamryE12,Shargel03,RePEc}.
A vast body of these works focuses on global robustness of graphs at large.

With respect to research on local robustness, 
Trajanovski \emph{et al.} aim to spot critical regions in a graph the destruction of which would cause the biggest harm to the network \cite{conf/infocom/TrajanovskiKM13}.
Similar works aim to identify the critical nodes and links of a network \cite{conf/incos/FujimuraM10,PhysRevE71015103,journals/ton/ShenNXT13,journals/coap/SummaGL12}.
These works try to spot vulnerability points in the network, whereas our objective is somewhat orthogonal: identify robust regions. Closest to ours,
Andersen {\em et al.} consider spectral radius as an objective criterion
 and propose algorithms for identifying small robust subgraphs with large spectral radius \cite{journals/jucs/AndersenC07}. 

While having major distinctions as we illustrated in this work, robust subgraphs are related to dense subgraphs, which have been studied extensively.
Finding the largest clique in
 a graph, well-known to be \textbf{NP}-complete \cite{gareyjohnson}, is also shown to be hard to approximate \cite{conf/focs/Hastad96}. 
 
 A relaxation of the clique problem is the densest subgraph problem. 
 Goldberg \cite{Goldberg84} and Charikar \cite{conf/approx/Charikar00} designed exact poly-time and $\frac{1}{2}$-approximate linear-time solutions to this problem, respectively, where density is defined as the average degree. This problem is shown to become {\bf NP}-hard when the size of the subgraph is restricted \cite{journals/dam/AsahiroHI02}.
 Most recently, Tsourakakis \textit{et al.} \cite{conf/kdd/TsourakakisBGGT13} also proposed fast heuristic solutions, 
 where they define density as edge surplus; the difference between number of edges and $\alpha$ fraction of maximum edges, for user-specified constant $\alpha>0$. Likewise,  Pei {\em et al.} study detecting quasi-cliques in multi-graphs \cite{conf/kdd/PeiJZ05}.
 Other definitions include $k$-cores, $k$-plexes, and $k$-clubs, etc. \cite{series/ads/LeeRJA10}. 
 
 Dense subgraph discovery is related to finding clusters in graphs, however with major distinctions.
 Most importantly, dense subgraph discovery has to do with absolute density where there exists a preset threshold for what is sufficiently dense. On the other hand, graph clustering concerns with relative density measures where density of one region is compared to another. Moreover, not all clustering objectives are based on density and not all types of dense subgraphs can be found by clustering algorithms \cite{series/ads/LeeRJA10}.
 
 In summary, while similarities among them exist, discovery of critical regions, robust subgraphs, cliques, densest subgraphs, and clusters are substantially distinct graph mining problems, for which different algorithms can be applied. To the best of our knowledge, our work is the first to consider identifying robust local subgraphs in large graphs.

\hide{

In graph mining research, finding and analyzing dense subgraphs is an important topic. It not only has significant meaning in theory but also is used to explain many real world phenomena. For example, from theoretical point of view, Motzkin and Straus~\cite{motzkin1965maxima} proved that solving a dense subgraph problem on an unweighted graph is equivalent to finding the maxima of a quadratic function on the simplex. 

From practical point of view, 
dense subgraphs find applications in several fields. In finance and economics, dense components in a market graph, where two financial instruments (e.g., bank, stock) that are correlated in their behaviors (e.g., loan transfers, price change over time) are connected, represent entities that are highly correlated and help understand the market dynamics, predict behaviors of entities, and indicate strength and robustness of the market \cite{journals/cor/BoginskiBP06}. Biology also benefit from dense subgraph discovery substantially. 
Dense components correspond to functional units of proteins in protein-protein interaction networks, 
 co-expression genes in gene regulation networks~\cite{junker2008analysis}, and food dependencies in ecological networks \cite{0470041447}. 
Finally, dense subgraphs relate to 
communities in social networks~\cite{newman2004detecting} such as phone-call graphs \cite{Abello99onmaximum}. We elaborate on the relationships between graph clustering and dense subgraph discovery below.

\paragraph{Cliques}

A clique is a subset of vertices that are all connected to
each other.
Finding the largest clique in
a graph, is a problem well-known to be \textbf{NP}-complete \cite{gareyjohnson}, and hard to approximate \cite{conf/focs/Hastad96}. 

\paragraph{Densest Subgraphs} 
Dense component discovery can be viewed as relaxation of clique discovery.
This relaxation has been done in various ways in prior works where density of a subgraph is defined differently.
 Abello {\em et al.} \cite{conf/latin/AbelloRS02} defined $\gamma$-dense quasi-cliques, or $\gamma$-quasi-cliques, 
as subgraphs that have at least $\gamma$ fraction of edges that a clique of the same size would have.
Another version of the quasi-clique problem asks for subgraphs in which all vertices connect to at least $\gamma$ fraction of the other vertices in the subgraph \cite{Jiang2009}. Note that this latter version is stricter in subgraph topology, since a solution for it is also a $\gamma$-quasi-clique but not vice-versa. Goldberg \cite{Goldberg84} and Charikar \cite{conf/approx/Charikar00} proposed exact poly-time and $\frac{1}{2}$-approximate linear-time solutions to the densest subgraph problem, respectively, where density is defined as the average degree. This problem is shown to become {\bf NP}-hard when the size of the subgraph is restricted \cite{journals/dam/AsahiroHI02}. 
Most recently, Tsourakakis \textit{et al.} \cite{conf/kdd/TsourakakisBGGT13} proposed fast heuristic solutions for the densest subgraph problem where they define density as edge surplus; the difference between number of edges and $\alpha$ fraction of edges a clique would have for user-specified constant $\alpha>0$. 
Other definitions include $k$-cores, $k$-plexes, and $k$-clubs \cite{series/ads/LeeRJA10}.


%

\paragraph{Graph Clustering}
Dense subgraph discovery is related to finding clusters in graphs, however with major distinctions.
Most importantly, dense subgraph discovery has to do with absolute density where there exists a preset threshold for what is sufficiently dense. On the other hand, graph clustering concerns with relative density measures where density of one region is compared to another. Moreover, not all clustering objectives are based on density and not all types of dense subgraphs can be found by clustering algorithms \cite{series/ads/LeeRJA10}.

\paragraph{Robust Subgraphs}

Andersen {\em et al.} study spectral versions of the densest subgraph problem and propose algorithms for identifying small subgraphs with large spectral radius \cite{journals/jucs/AndersenC07}. 
}

\vspace{0.05in}
\section{Conclusion}
\label{sec:conclusion}

We introduced the \rls~of finding the most robust local subgraph of a given size in large graphs, as well as its three practical variants. While our work bears similarity to densest subgraph mining, it differs from it in its objective; robustness emphasizes subgraph topology more than edge density. We showed that our problem is {\bf NP}-hard and that it does not exhibit semi-heredity or subgraph monotonicity properties. We designed two heuristic algorithms based on top-down and bottom-up search strategies, and showed how we can adapt them to address the problem variants. 
We found that our bottom-up strategy provides consistently superior results, scales linearly with input graph size, and finds subgraphs with significant robustness.
Experiments on synthetic and real graphs showed that our subgraphs are of higher robustness than densest subgraphs even at lower densities, which illustrates the novelty of our problem setting.  

Our research sets off several future directions, including the hardness analysis for the \rgs,  exploration of new  robustness measures with desirable properties, and the design of efficient algorithms for those new objectives.

\bibliographystyle{abbrv}
\bibliography{BIB/robustness,BIB/refsmeas}

\begin{thebibliography}{10}

\bibitem{albert2000error}
R.~Albert, H.~Jeong, and A.-L. Barabasi.
\newblock Error and attack tolerance of complex networks.
\newblock {\em Nature}, 406(6794), 2000.

\bibitem{journals/jucs/AndersenC07}
R.~Andersen and S.~M. Cioaba.
\newblock Spectral densest subgraph and independence number of a graph.
\newblock {\em J. UCS}, 13(11), 2007.

\bibitem{journals/dam/AsahiroHI02}
Y.~Asahiro, R.~Hassin, and K.~Iwama.
\newblock Complexity of finding dense subgraphs.
\newblock {\em Disc. Appl. Math.}, 121(1-3):15--26, 2002.

\bibitem{journals/jal/AsahiroITT00}
Y.~Asahiro, K.~Iwama, H.~Tamaki, and T.~Tokuyama.
\newblock Greedily finding a dense subgraph.
\newblock {\em J. Algorithms}, 34(2), 2000.

\bibitem{Beygelzimer2005}
A.~Beygelzimer, G.~Grinstein, R.~Linsker, and I.~Rish.
\newblock Improving network robustness by edge modification.
\newblock {\em Physica A: Stat. Mech. and its Appl.}, 357(3-4):593--612, 2005.

\bibitem{callaway2000nra}
D.~S. Callaway, M.~E.~J. Newman, S.~H. Strogatz, and D.~J. Watts.
\newblock {Network Robustness and Fragility: Percolation on Random Graphs}.
\newblock {\em Phys. Rev. Let.}, 85(25):5468--5471, 2000.

\bibitem{ChanSDM14}
H.~Chan, L.~Akoglu, and H.~Tong.
\newblock Make it or break it: Manipulating robustness in large networks.
\newblock In {\em SDM}, pages 325--333, 2014.

\bibitem{conf/approx/Charikar00}
M.~Charikar.
\newblock Greedy approximation algorithms for finding dense components in a
  graph.
\newblock In {\em APPROX}, 2000.

\bibitem{ChungLu03}
F.~R.~K. Chung and L.~Lu.
\newblock The average distance in a random graph with given expected degrees.
\newblock 1(1), 2003.

\bibitem{citeulike2352886}
R.~Cohen, K.~Erez, D.~B. Avraham, and S.~Havlin.
\newblock {Breakdown of the {Internet} under Intentional Attack}.
\newblock {\em Physical Review Letters}, 86(16):3682--3685, Apr. 2001.

\bibitem{journals/corr/EllensK13}
W.~Ellens and R.~E. Kooij.
\newblock Graph measures and network robustness.
\newblock {\em CoRR}, abs/1311.5064, 2013.

\bibitem{journals/bioinformatics/Estrada02}
E.~Estrada.
\newblock Characterization of the folding degree of proteins.
\newblock {\em Bioinformatics}, 18(5):697--704, 2002.

\bibitem{expansion2006}
E.~Estrada.
\newblock Network robustness to targeted attacks: {The} interplay of
  expansibility and degree distribution.
\newblock {\em The Euro. Phys. J. B}, 52(4):563--574, 2006.

\bibitem{journals/corr/abs-1109-2950}
E.~Estrada, N.~Hatano, and M.~Benzi.
\newblock The physics of communicability in complex networks.
\newblock {\em CoRR}, abs/1109.2950, 2011.

\bibitem{conf/sigcomm/FaloutsosFF99}
M.~Faloutsos, P.~Faloutsos, and C.~Faloutsos.
\newblock On power-law relationships of the internet topology.
\newblock In {\em SIGCOMM}, 1999.

\bibitem{Feo95greedyrandomized}
T.~A. Feo and M.~G. Resende.
\newblock Greedy randomized adaptive search procedures.
\newblock {\em J. of Optimization}, 6:109--133, 1995.

\bibitem{FrankFrisch1970}
H.~Frank and I.~Frisch.
\newblock {Analysis and Design of Survivable Networks}.
\newblock {\em IEEE Trans. on Comm. Tech.}, 18(5), 1970.

\bibitem{conf/incos/FujimuraM10}
T.~Fujimura and H.~Miwa.
\newblock Critical links detection to maintain small diameter against link
  failures.
\newblock In {\em INCoS}, pages 339--343. IEEE, 2010.

\bibitem{gareyjohnson}
M.~Garey and D.~Johnson.
\newblock {\em {C}omputers and {I}ntractability - {A} guide to the {T}heory of
  {NP}-{C}ompleteness}.
\newblock Freeman, 1979.

\bibitem{journals/telsys/GhamryE12}
W.~K. Ghamry and K.~M.~F. Elsayed.
\newblock Network design methods for mitigation of intentional attacks in
  scale-free networks.
\newblock {\em Telecom. Systems}, 49(3):313--327, 2012.

\bibitem{Goldberg84}
A.~V. Goldberg.
\newblock Finding a maximum density subgraph.
\newblock Technical Report CSD-84-171, UC Berkeley, 1984.

\bibitem{conf/focs/Hastad96}
J.~Hastad.
\newblock Clique is hard to approximate within $n^{(1-\epsilon)}$.
\newblock In {\em FOCS}, pages 627--636. IEEE Computer Society, 1996.

\bibitem{Holme2002}
P.~Holme, B.~J. Kim, C.~N. Yoon, and S.~K. Han.
\newblock Attack vulnerability of complex networks.
\newblock {\em Phy. R. E}, 65(5), 2002.

\bibitem{journals/jcss/JohnsonPY88}
D.~S. Johnson, C.~H. Papadimitriou, and M.~Yannakakis.
\newblock How easy is local search?
\newblock {\em J. Comp. Sys. Sci.}, 37(1), 1988.

\bibitem{PhysRevE71015103}
V.~Latora and M.~Marchiori.
\newblock Vulnerability and protection of infrastructure networks.
\newblock {\em Phys. Rev. E}, 71:015103, 2005.

\bibitem{series/ads/LeeRJA10}
V.~E. Lee, N.~Ruan, R.~Jin, and C.~C. Aggarwal.
\newblock A survey of algorithms for dense subgraph discovery.
\newblock In {\em Managing and Mining Graph Data}. Springer, 2010.

\bibitem{MalliarosMF12}
F.~D. Malliaros, V.~Megalooikonomou, and C.~Faloutsos.
\newblock Fast robustness estimation in large social graphs: Communities and
  anomaly detection.
\newblock In {\em SDM}, pages 942--953, 2012.

\bibitem{journals/dam/PattilloVBB13}
J.~Pattillo, A.~Veremyev, S.~Butenko, and V.~Boginski.
\newblock On the maximum quasi-clique problem.
\newblock {\em Discrete Applied Mathematics}, 161(1-2):244--257, 2013.

\bibitem{RePEc}
G.~Paul, T.~Tanizawa, S.~Havlin, and H.~Stanley.
\newblock Optimization of robustness of complex networks.
\newblock {\em The Eur. Phys. J. B}, 38(2):187--191, 2004.

\bibitem{conf/kdd/PeiJZ05}
J.~Pei, D.~Jiang, and A.~Zhang.
\newblock On mining cross-graph quasi-cliques.
\newblock In {\em KDD}, pages 228--238, 2005.

\bibitem{Shargel03}
B.~Shargel, H.~Sayama, I.~R. Epstein, and Y.~Bar-Yam.
\newblock Optimization of robustness and connectivity in complex networks.
\newblock {\em Phys Rev Lett}, 90(6):068701, 2003.

\bibitem{journals/ton/ShenNXT13}
Y.~Shen, N.~P. Nguyen, Y.~Xuan, and M.~T. Thai.
\newblock On the discovery of critical links and nodes for assessing network
  vulnerability.
\newblock {\em IEEE/ACM Trans. Netw.}, 21(3), 2013.

\bibitem{MatrixPerturb}
G.~W. Stewart and J.-G. Sun.
\newblock {\em Matrix Perturbation Theory}.
\newblock Academic Press, 1990.

\bibitem{journals/coap/SummaGL12}
M.~D. Summa, A.~Grosso, and M.~Locatelli.
\newblock Branch and cut algorithms for detecting critical nodes in undirected
  graphs.
\newblock {\em Comp. Opt. and Appl.}, 53(3):649--680, 2012.

\bibitem{journals/amc/SydneySG13}
A.~Sydney, C.~M. Scoglio, and D.~Gruenbacher.
\newblock Optimizing algebraic connectivity by edge rewiring.
\newblock {\em Applied Mathematics and Computation}, 219(10), 2013.

\bibitem{conf/cikm/TongPEFF12}
H.~Tong, B.~A. Prakash, T.~Eliassi-Rad, M.~Faloutsos, and C.~Faloutsos.
\newblock Gelling, and melting, large graphs by edge manipulation.
\newblock In {\em CIKM}, pages 245--254, 2012.

\bibitem{conf/infocom/TrajanovskiKM13}
S.~Trajanovski, F.~A. Kuipers, and P.~V. Mieghem.
\newblock Finding critical regions in a network.
\newblock In {\em INFOCOM}, 2013.

\bibitem{conf/icdm/Tsourakakis08}
C.~E. Tsourakakis.
\newblock Fast counting of triangles in large real networks without counting:
  Algorithms and laws.
\newblock In {\em ICDM}, pages 608--617. IEEE Computer Society, 2008.

\bibitem{conf/kdd/TsourakakisBGGT13}
C.~E. Tsourakakis, F.~Bonchi, A.~Gionis, F.~Gullo, and M.~A. Tsiarli.
\newblock Denser than the densest subgraph: extracting optimal quasi-cliques
  with quality guarantees.
\newblock In {\em KDD}, 2013.

\bibitem{Watts2003}
D.~Watts.
\newblock {Security and vulnerability in electric power systems}.
\newblock {\em 35th N. American Power Symp.}, pages 559--566, 2003.

\bibitem{WUJun78902}
J.~Wu, B.~Mauricio, Y.-J. Tan, and H.-Z. Deng.
\newblock Natural connectivity of complex networks.
\newblock {\em Chin. Phys. Lett.}, 27(7), 2010.

\bibitem{journals/corr/abs-1203-2982}
A.~Zeng and W.~Liu.
\newblock Enhancing network robustness for malicious attacks.
\newblock {\em CoRR}, abs/1203.2982, 2012.

\end{thebibliography}

\newpage
\clearpage
\section*{\Large Appendix}
\label{sec:appendix}


\setcounter{section}{0}

\renewcommand\thesection{\Alph{section}}

\vspace{0.05in}

\section{NP-Hardness Proof of \rls} 
\label{sec:npproof}

The decision version of the \rls~is as follows.

\vspace{0.05in}
\noindent
{\bf P1.} (\textsf{robust $s$-subgraph problem} \rls) Is there a subgraph $S$ in graph $G$ with $|S|=s$ nodes and robustness $\bar{\lambda}(S) \geq \alpha$, for some $\alpha \geq 0$?

In order to show that {\bf P1} is {\bf NP}-Hard, we reduce from the NP-Complete $s$-clique problem {\bf P2} \cite{gareyjohnson}.

\noindent
{\bf P2.} (\textsf{$s$-clique problem} {\sc CL}) Does graph $G$ contain a clique of size $s$?

\begin{proof} (of {\sc Theorem} \ref{th:hardness})

It is easy to see that P1 is in NP, since given a graph $G$ we can guess the subgraph with $s$ nodes and compute its robustness in polynomial time.	

In the reduction, 
we convert the instances as follows. An instance of {\sc CL} is a graph $G=(V,E)$ and an integer $s$. We pass $G$, $s$, and $\alpha = \bar{\lambda}(C_s)$ to \rls, where $C_s$ is a clique of size $s$.
We show that a {\em yes} instance of {\sc CL} maps to a {\em yes} instance of \rls, and vice versa. 

First assume $C$ is a {\em yes} instance of {\sc CL}, i.e., there exists a clique of size $s$ in $G$. Clearly the same is also a {\em yes} instance of \rls~as $\bar{\lambda}(C_s) \geq \alpha$.

Next assume $S$ is a {\em yes} instance of \rls, thus $\bar{\lambda}(S) \geq \bar{\lambda}(C_s)$. The proof is by contradiction. Assume $S$ is a subgraph with $s$ nodes that is {\em not} a clique. As such, it should have one or more missing edges from $C_s$. 
Let us denote by $W_k = \text{trace}(\mathbf{A}_{C_s}^k)$ the number of closed walks of length $k$ in $C_s$.
Deleting an edge from $C_s$, $W_k$ will certainly not increase, and in some cases (e.g., for $k = 2$) will strictly decrease. As such, any $s$-subgraph $S'$ of $C_s$ with missing edges will have $\bar{\lambda}(S') < \bar{\lambda}(C_s)$, which is a contradiction to our assumption that $S$ is a {\em yes} instance of the \rls. Thus, $S$ should be an $s$-clique and also a {\em yes} instance of {\sc CL}. $\square$
\end{proof}

\section{Properties of Robust Subgraphs} 
\label{sec:properties}

In this section we study two properties of the \rls, namely semi-heredity (a.k.a. quasi-inheritance) and subgraph monotonicity. 
Analysis shows that our objective formulation does not exhibit neither of these properties.

\noindent
\subsection{\bf Semi-heredity}: It is easy to identify $\alpha$-robust graphs  (i.e., $\bar{\lambda}=\alpha$) that contain subsets of nodes that induce subgraphs with robustness less than $\alpha$. As such, robust subgraphs do not display heredity.
Here, we study a weaker version of heredity called semi-heredity or quasi-inheritance.

\begin{definition}[Semi-heredity]
Given any graph $G = (V, E)$ satisfying a property $p$, if there exists {\em some} $v\in V$ such that 
$G-v \equiv G[V\backslash \{v\}]$ also has property $p$, $p$ is called a semi-hereditary property.
\end{definition}
%

\begin{proof} (of {\sc Theorem \ref{th:semiheredity}})

The proof is by counter example. In particular, robustness of cliques is not semi-hereditary. Without loss of generality, let $C_k$ be a $k$-clique. Then, $\bar{\lambda}(C_k) = \ln \frac{1}{k} (e^{k-1} + (k-1)\frac{1}{e})$. \textit{Any} subgraph of $C_k$ with $k-1$ nodes is also a clique having strictly lower robustness, for $k>1$, i.e.,
\begin{align}
\vspace{-0.1in}
\frac{1}{k-1} (e^{k-2} + (k-2)\frac{1}{e}) & \leq \frac{1}{k} (e^{k-1} + (k-1)\frac{1}{e}) \nonumber \\
k e^{k-2} + \frac{k(k-2)}{e} & \leq (k-1)e^{(k-1)} + \frac{(k-1)^2}{e} \nonumber \\
k e^{k-1} + k^2-2k & \leq (k-1)e^{k} + (k^2 - 2k + 1) \nonumber \\
k e^{k-1} & \leq (k-1)e^{k} + 1 \nonumber 
\end{align}
where the inequality is sharp only for $k=1$. Thus, for $\alpha = \bar{\lambda}(C_k)$, there exists {\em no} $v$ such that $C_k-v$ is at least $\alpha$-robust.
$\square$
\end{proof}

\subsection{\bf Subgraph monotonicity}:
As we defined in \S\ref{subsec:robst}, our robustness measure can be written in terms of subgraph centrality as
$\bar{\lambda}(G) = \log (\frac{1}{n} S(G))$.

As $S(G)$ is the total number of weighted closed walks, $\bar{\lambda}$ is strictly monotonic with respect to edge additions and deletions.
However, monotonicity is not directly obvious for node modifications due to the $\frac{1}{n}$ factor in the definition, which changes with the graph size. 

\begin{definition}[Subgraph monotonicity]
An objective function (in our case, robustness) $R$  is {\em subgraph monotonic} if for any subgraph $g = (V', E')$ of $G = (V, E)$, $V' \subseteq V$ and $E' \subseteq E$, $R(g) \leq R(G)$.
\end{definition}
%
%

\begin{figure}[!t]
\centering
\includegraphics[width=0.25\textwidth]{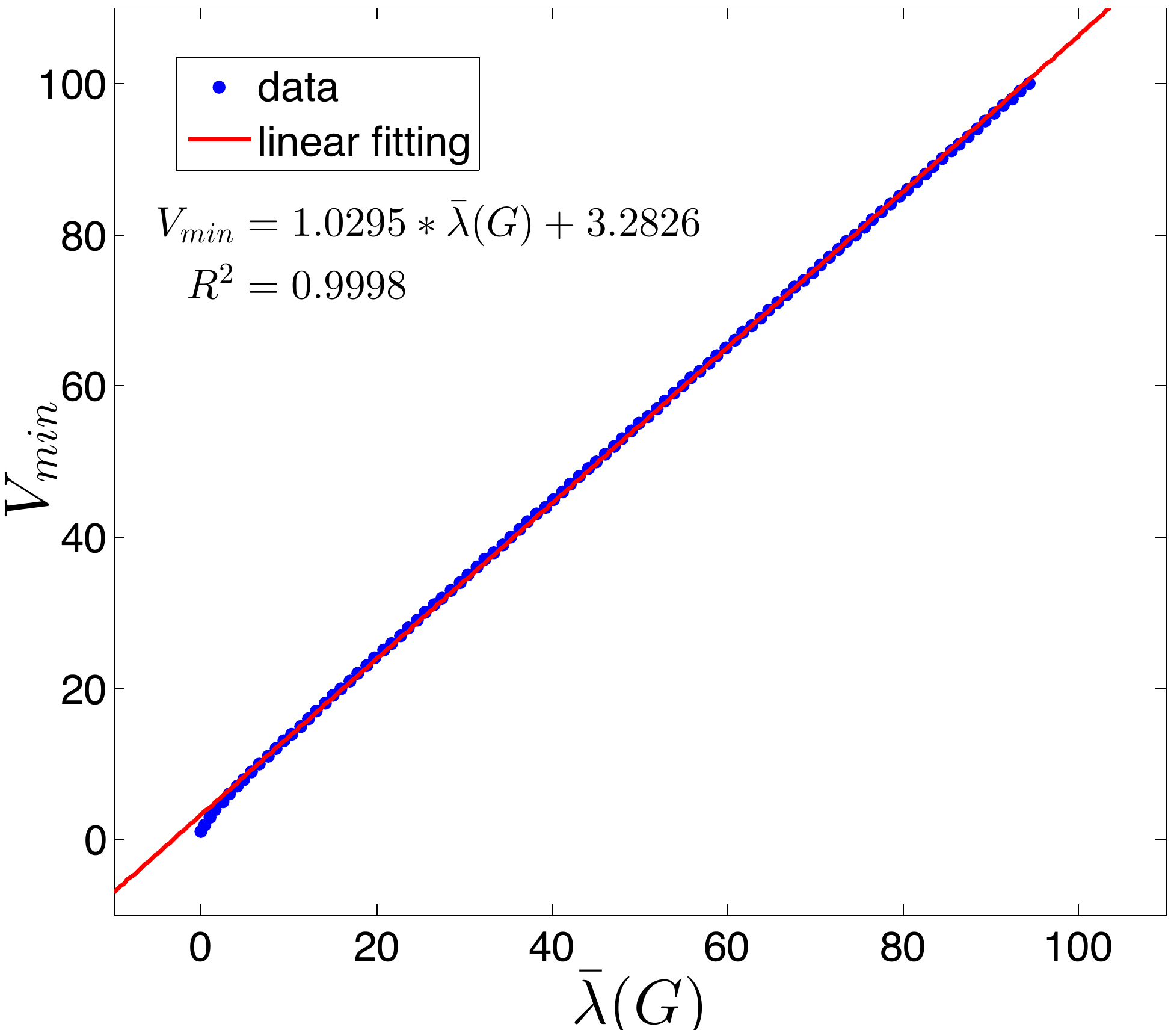}
\vspace{-0.15in}
\caption{\small Relation between $\bar{\lambda}(G)$ and $V_{\min}$} 
\label{fig:robustness2clique}
\end{figure}

\begin{proof} (of {\sc Theorem \ref{th:monoton}})

Assume we start with any graph $G$ with robustness $\bar{\lambda}(G)$. Next, we want to find a graph $S$ with as large robustness as $\bar{\lambda}(G)$ but that contains the minimum possible number of nodes $V_{\min}$.
Such a smallest subgraph is in fact a clique, with the largest eigenvalue ($V_{\min}-1$) and the rest of the $(V_{\min}-1)$ eigenvalues equal to $-1$.\footnote{\scriptsize Any subgraph $g(C)$ of a $k$-clique $C$ has strictly lower robustness $\bar{\lambda}$. This is true when $g(C)$ also contains $k$ nodes, due to monotonicity of $S(G)$ to edge removals (see Appendix \ref{sec:npproof}). Any smaller clique has strictly lower robustness, see proof for semi-heredity.
} To obtain the exact value of $V_{\min}$, we need to solve the following

$$
\bar{\lambda}(G) = \log \left[\frac{1}{V_{\min}}\left(e^{V_{\min}-1}+(V_{\min}-1) e^{-1}\right)\right] 
$$

which, however, is a transcendental equation and is often solved using numerical methods. To obtain a solution quickly, we calculate a linear regression over $(\bar{\lambda}(G), V_{\min})$ samples. We show a sample simulation in Figure \ref{fig:robustness2clique} for $V_{\min}$ 1 to 100 where the regression equation is

\begin{equation*}
\label{equ:vmin}
V_{\min} = 1.0295*\bar{\lambda}(G) + 3.2826
\end{equation*}

Irrespective of how one computes $V_{\min}$, we next construct a new graph $G' = G\cup S'$ in which $G$ is the original graph with $n$ nodes and $S'$ is a clique of size $V_{\min}+1$. Let $\lambda = \bar{\lambda}(G)$ and 
$\lambda' = \bar{\lambda}(S')$, and as such, $\lambda < \lambda'$.
Then, we can write the robustness of $G'$ as

$$
\bar{\lambda}(G') = \ln \frac{ne^{\lambda} + (V_{\min}+1)e^{\lambda'}}{n+V_{\min}+1}
$$
$$ \;\;\;\;\;\;\; \;\;\;\;\;\;\;\;\;\;\;\;< 
\ln \frac{ne^{\lambda'} + (V_{\min}+1)e^{\lambda'}}{n+V_{\min}+1} = \lambda'
$$

which shows that $S'$, which is a subgraph of $G'$, has strictly larger robustness than the original graph. This construction shows that $\bar{\lambda}$ is not subgraph monotonic. 
$\square$
\end{proof}

\section{\rls~Variants}
\label{sec:variants}

In this section, we describe three practical variants of the original \rls~problem.

Given that robustness $\bar{\lambda}$ is not subgraph-monotonic, it is natural to consider the problem of 
 finding the subgraph with the maximum overall robustness in the graph (without any restriction on its size).
We call this first variant the robust global subgraph problem or the \rgs. 
Note that {\sc RGS} is not necessarily the full graph.

\begin{problem}[\rgs]
Given a graph $G = (V,E)$, find a subgraph $S^* \subseteq V$ such that

$$
f(S^*) = \max_{S \subseteq V} f(S) \;.
$$

$S^*$  is referred as the \textsf{\em most robust subgraph}.
\end{problem}

Another variant involves finding the top $k$ most robust $s$-subgraphs in a graph, which we call the \krls.

\begin{problem}[\krls]
\label{prob:krls}
Given a graph $G = (V,E)$, and two integers $s$ and $k$, find $k$ subgraphs $\mathcal{S} = S^*_1, \ldots, S^*_k$, each of size $|S^*_i|=s$, $1\leq i \leq k$ such that

$$
f(S^*_1) \geq f(S^*_2) \geq \ldots \geq f(S^*_k)
 \geq f(S), \;\; \forall S \subseteq V, |S|=s \;.
$$

$\mathcal{S}$  is referred as the \textsf{\em top-$k$ most robust $s$-subgraphs}.
\end{problem}

In the third and final variant, the goal is to find the most robust $s$-subgraph that contains a set of {\em user-given} seed nodes, called the \seedrls.

\begin{problem}[\seedrls]
\label{prob:seedrls}
Given a graph $G = (V,E)$, an integer $s$, and a set of seed nodes $U$, $|U| \leq s$, find a subgraph $U \subseteq S^* \subseteq V$ of size $|S^*|=s$ such that

$$
f(S^*) = \max_{U \subseteq S \subseteq V, |S|=s}  f(S) \;.
$$

$S^*$  is referred as the \textsf{\em most robust seeded $s$-subgraph}.
\end{problem}

It is easy to see that when $k=1$ and $U = \emptyset$, the \krls~and the \seedrls~respectively reduce to the \rls, and thus can easily be shown to be also {\bf NP}-hard. A formal proof of hardness for the \rgs, however, is nontrivial and remains an interesting open problem for future research.

\section{Updating the Eigen-pairs} 
\label{ssec:pair}

\subsection{Updating the Eigen-values}
\label{ssec:eig}

\begin{proof} (of {{\sc Lemma} \ref{lemma:eigenvalueupdate}})

Using the relation $\mathbf{A} \mathbf{u_j} = \lambda_j \mathbf{u_j}$ between the eigenvalues and eigenvectors, we can write
the updated relation as
$$
(\mathbf{A} + \Delta \mathbf{A}) (\mathbf{u_j} + \Delta \mathbf{u_j}) = (\lambda_j + \Delta \lambda_j) (\mathbf{u_j} + \Delta \mathbf{u_j})
$$
Expanding the above, we get 
\begin{eqnarray}
\label{equ0}
\mathbf{A} \mathbf{u_j} +  \Delta \mathbf{A} \mathbf{u_j} + \mathbf{A}   \Delta \mathbf{u_j} + \Delta \mathbf{A}  \Delta \mathbf{u_j} \nonumber\\
= \lambda_j \mathbf{u_j} + \Delta \lambda_j \mathbf{u_j} + \lambda_j \Delta \mathbf{u_j} + \Delta \lambda_j \Delta \mathbf{u_j} 
\end{eqnarray}

By concentrating on first-order approximation, we assume that all high-order perturbation terms are negligible, including  $\Delta \mathbf{A}  \Delta \mathbf{u_j}$ and
$\Delta \lambda_j \Delta \mathbf{u_j}$. Further, by using the fact that $\mathbf{A} \mathbf{u_j}  = \lambda_j \mathbf{u_j}$ (i.e., canceling these terms) we obtain

\beq
\label{interm}
 \Delta \mathbf{A} \mathbf{u_j} + \mathbf{A}   \Delta \mathbf{u_j}  = \Delta \lambda_j \mathbf{u_j} + \lambda_j \Delta \mathbf{u_j}
\eeq

Next we multiply both sides by $\mathbf{u_j}'$ and by symmetry of \A~and orthonormal property of its eigenvectors we get Equ. (\ref{updatel}), which concludes the proof. \end{proof}

\subsection{Updating the Eigen-vectors}
\label{ssec:eigv}

\begin{proof} (of {{\sc Lemma} \ref{lemma:eigenvectorupdate}})

Using the orthogonality property of the eigenvectors, we can write the change $\Delta$\uj~of eigenvector \uj~as a linear combination of the original eigenvectors:

\beq
\label{equx}
\Delta \mathbf{u_j} = \sum\limits_{i=1}^n \alpha_{ij} \mathbf{u_i}
\eeq
where $ \alpha_{ij}$'s are small constants that we aim to determine.

Using Equ. (\ref{equx}) in Equ. (\ref{interm})
we obtain
$$
 \Delta \mathbf{A} \mathbf{u_j} + \mathbf{A}   \sum\limits_{i=1}^n \alpha_{ij} \mathbf{u_i}  = \Delta \lambda_j \mathbf{u_j} + \lambda_j \sum\limits_{i=1}^n \alpha_{ij} \mathbf{u_i}
$$
which is equivalent to
$$
 \Delta \mathbf{A} \mathbf{u_j} + \sum\limits_{i=1}^n \lambda_i \alpha_{ij} \mathbf{u_i}  = \Delta \lambda_j \mathbf{u_j} + \lambda_j \sum\limits_{i=1}^n \alpha_{ij} \mathbf{u_i}
$$
Multiplying both sides of the above by $\mathbf{u_k}'$, $k\neq j$, we get
$$
\mathbf{u_k}' \Delta \mathbf{A} \mathbf{u_j}  +   \lambda_k \alpha_{kj} =  \lambda_j \alpha_{kj}
$$
Therefore,

\beq
\label{alphas}
\alpha_{kj} = \frac{\mathbf{u_k}' \Delta \mathbf{A} \mathbf{u_j} }{\lambda_j - \lambda_k}
\eeq
for $k\neq j$. To obtain $\alpha_{jj}$ we use the following derivation.
\begin{align}
\nonumber
                    \tilde{\mathbf{u_j}}' \tilde{\mathbf{u_j}} = 1
& \Rightarrow (\mathbf{u_j} + \Delta \mathbf{u_j})'(\mathbf{u_j} + \Delta \mathbf{u_j}) = 1\\ \nonumber
&\Rightarrow  1 + 2 \mathbf{u_j}' \Delta \mathbf{u_j} + \| \Delta \mathbf{u_j} \|^2 = 1
\end{align}
After we discard the high-order term, and substitute $\Delta \mathbf{u_j}$ with Equ. (\ref{equx}) we get $1 + 2 \alpha_{jj} = 1 \Rightarrow  \alpha_{jj} =0$.

We note that for a slightly better approximation, one can choose not to ignore the high-order term which is equal to $\| \Delta \mathbf{u_j} \|^2 =  \sum\limits_{i=1}^n \alpha_{ij}^2$.
Thus, one can compute $\alpha_{jj}$ as

\begin{align}
\nonumber
& 1+ 2 \alpha_{jj} + \sum\limits_{i=1}^n \alpha_{ij}^2 = 1 \\ \nonumber
& \Rightarrow  1+ 2 \alpha_{jj} + \alpha_{jj}^2 +\sum\limits_{{i=1, i\neq j}}^n  \alpha_{ij}^2 = 1\\ \nonumber
&\Rightarrow  (1+\alpha_{jj})^2 + \sum\limits_{{i=1, i\neq j}}^n  \alpha_{ij}^2 = 1 \\ \nonumber
& \Rightarrow   \alpha_{jj} = \sqrt{1-\sum\limits_{{i=1, i\neq j}}^n \alpha_{ij}^2} - 1
\end{align}

Using the $\alpha_{ij}$'s as given by Equ. (\ref{alphas}) and $\alpha_{jj}=0$, we can see that $\Delta \mathbf{u_j}$ in Equ. (\ref{equx}) is equal to Equ. (\ref{updateu}). 
\end{proof}

\vspace{2in}

\hide{
\noindent
{\bf Higher order approximation for eigenvectors.} 

In this section, we derive a `better' method to update the $t$ eigenvectors 
(keep in mind that we are still using 
 Equ.~(\ref{updatel}) to estimate $\Delta \lambda_j (j=1,...,t)$).

In \textit{Lemma \ref{lemma:eigenvectorupdate}}, 
we establish that we can update $\Delta \mathbf{u_j}$ via
Equ. (\ref{updateu}) quite efficiently. 
In the proof of \textit{Lemma \ref{lemma:eigenvectorupdate}}, 
we write the change $\Delta \mathbf{u_j}$ as a linear combination 
of the original eigenvectors:  
\beq
\label{equx2}
\Delta \mathbf{u_j} = \sum\limits_{i=1}^t \alpha_{ij} \mathbf{u_i}
\eeq
then substitute the expression into Equ. (\ref{interm}) and solve for the $\alpha_{ij}$ terms.

However, Equ. (\ref{interm}) is obtained from Equ. (\ref{equ0})
by removing high-order perturbation terms. 
It appears that if we plug Equ.~\eqref{equx2} into Equ.~\eqref{equ0}
and keep the higher order terms, we could provide a better estimation of the eigenvectors 
by selecting `better' $\alpha$ values.

Now, suppose that we plug Equ.~\eqref{equx2} into Equ.~\eqref{equ0} 
and multiply both sides by $\mat{u}_k' \;\; (k=1,...,t)$, $k \neq j$.  Then, for a given $\mat{u}_j \;\; (j=1,...,t)$ we have 
\beq
\mat{X}(k,j) + (\lambda_k - \lambda_j - \Delta \lambda_j)\alpha_{k,j} + \sum_{i=1}^t \mat{X}(k,i)\alpha_{i,j} = 0
\label{eq:2ndu}
\eeq
\noindent where $\mat{X}(k,i) = \mat{u}_k' \Delta \mat{A} \mat{u}_i \;\; (k,i=1,..., t)$.

For simplicity, we let $\mat{\alpha}_j = [\alpha_{1,j},...,\alpha_{k,j}]'$ and $\mat{D} = \textrm{diag}(\lambda_j + \Delta \lambda_j -\lambda_k)$ for $k=1,...,t$.
In this matrix form, we have the following linear system for $\mat{\alpha}_j$:
\[(\mat{D} - \mat{X})\mat{\alpha}_j = \mat{X}(:,j)\]

Solving for $\alpha_j$, we have 
\begin{center}
$\mat{\alpha}_j = (\mat{D} - \mat{X})^{-1}\mat{X}(:,j)$ 
\end{center}

We can see that the above new formula includes the 
original estimation in Equ.~\eqref{alphas} as a special case by dropping $\Delta \lambda_j$ and $\mat X$.
Theoretically, this will give us a better estimation, 
and help us get around the multiplicity issue of the eigenvalues.
However, this method of update is more expensive; the new formula increases the time complexity by an additional $t^4$ 
where $t$ is the number of eigenvectors. 
Thus, Equ. (\ref{updateu}) is more appropriate in a large network setting.
}

\section{Greedy Top-down Search for \rls} 
\label{sec:greedyalg}

\subsection{\greedymrs~Algorithm}

See pseudo-code in Algorithm \ref{alg:greedy}.

\begin{algorithm}[h]
\caption{\greedymrs} \label{alg:greedy}
\begin{algorithmic}[1]
\REQUIRE Graph $G = (V,E)$, its adj. matrix \A, integer $s$
\ENSURE Subset of nodes $S^* \subseteq V$, $|S^*| = s$
\STATE Compute top $t$ eigen-pairs $(\lambda_j,\mathbf{u_j})$ of $\mathbf{A}$, $1 \leq j \leq t$
\STATE $S_n \leftarrow V$, $\bar{\lambda}(S_n) = \bar{\lambda}(G)$
\FOR{$z = n$ down to $s+1$}
\STATE Select node $\bar{i}$ out of $\forall i \in S_z$ that maximizes Equ. (\ref{picknode}) for top $t$ eigen-pairs of $G[S_z]$, i.e.
$$
f = \max\limits_{i\in S_z} \hspace{0.1in} c_1 \bigg(e^{-2 \mathbf{u_{i1}} \sum\limits_{v\in {\cal N}(i)} \mathbf{u_{v1}}} + \ldots + c_t e^{-2 \mathbf{u_{it}} \sum\limits_{v\in {\cal N}(i)} \mathbf{u_{vt}}} \bigg)
$$
where $c_1=e^{\lambda_1}$ and $c_j=e^{(\lambda_j-\lambda_1)}$ for $2\leq j\leq t$
\STATE $S_{z-1} := S_z \backslash \{\bar{i}\}$, $\bar{\lambda}(S_{z-1}) = \log \frac{f}{z-1}$
\STATE Update \A; $\mathbf{A}(:,\bar{i})=0$ and $\mathbf{A}(\bar{i},:)=0$

\IF {$z=\frac{n}{2}, \frac{n}{4}, \frac{n}{8}, \ldots$}
\STATE Compute top $t$ eigen-pairs $(\lambda_j,\mathbf{u_j})$ of $\mathbf{A}$, $1 \leq j \leq t$
\ELSE
\STATE Update top $t$ eigenvalues of \A~by Equ. (\ref{updatel})
\STATE Update  top $t$ eigenvectors of \A~by Equ. (\ref{updateu})
\ENDIF
\ENDFOR
\RETURN $S^* \leftarrow S_{z=s}$ 
\end{algorithmic}
\end{algorithm}

\subsection{Complexity analysis}
Algorithm \ref{alg:greedy} has three main components: 
(a) computing top $t$ eigenpairs (L1): $O(nt+mt+nt^2)$, 
(b) computing Equ. (\ref{picknode}) scores for all nodes using top $t$ eigen-pairs (L4): $O(mt)$  ($\sum_i d_it = t \sum_i d_i = 2mt$),
and (c) updating $t$ eigenvalues:  $O(d_it)$ (using  Equ. (\ref{picknode1}))   \& updating $t$ eigenvectors: $O(nt^2)$ (using  Equ. (\ref{updateu})) when a node $i$ is removed (L10 \& L11, respectively).

Step (a) is executed only once, whereas (b) and (c) are performed at every step for $n-s$, i.e., $O(n)$ steps in general for small constant $s$.
Performing (b) for all nodes at every iteration thus takes $O(mtn)$.
Moreover, performing (c) iteratively for all nodes requires 
$\sum_{i=1}^{n} d_it = t \sum_i d_i = 2mt$, i.e., $O(mt)$ for eigenvalues and $\sum_{i=k}^{n} it^2, O(t^2n^2)$ for eigenvectors. 
Therefore, the overall complexity becomes $O(\max(tmn, t^2n^2))$. 


As we no longer would have small perturbations to the adjacency matrix over many iterations,
updating the eigen-pairs at all steps would yield bad approximations. As such, we recompute the eigen-pairs at every $\frac{n}{2}, \frac{n}{4}, \frac{n}{8}, \ldots$ steps. Performing recomputes less frequently in early iterations is reasonable, as early nodes are likely the peripheral ones that do not affect the eigen-pairs much, for which updates would suffice. When perturbations accumulate over iterations and especially when we get closer to the solution, it becomes beneficial to recompute the eigen-pairs more frequently.

In fact, in a greedier version one can drop the eigen-pair updates (L10-11), so as to only perform  $O(\log n)$ recomputes (L8), in which case the complexity becomes
 $O(\max(tm\log n, t^2n\log n))$.


\vspace{-0.05in}
\subsection{Algorithm Variants}
\label{ssec:var}

To adapt our \greedymrs~algorithm for the \krls, we can find one subgraph at a time, remove all its nodes from the graph, and continue until we find $k$ subgraphs or end up with an empty graph. 
This way we generate  node-disjoint robust subgraphs. One can also create edge-disjoint subgraphs if instead of removing the nodes of each found subgraph, one removes the edges among the nodes of the subgraph.

For the \seedrls, we can condition to never remove nodes $u\in U$ that belong to the seed set. 

\greedymrs~algorithm is particularly suitable for the \rgs, where we can iterate for $z=n, \ldots, V_{\min}$\footnote{\scriptsize $V_{\min}$ denotes the minimum number of nodes a clique $C$ with robustness at least as large as the full graph's robustness would contain. Any subgraph of the clique $C$ has lower robustness (see Appendix \ref{sec:properties}) and hence would not qualify as the \textsf{most robust subgraph}.}, 
record the robustness $\bar{\lambda}(S_{z})$ for each subgraph at each step (Alg.\ref{alg:greedy} L5), and return the subgraph  with the maximum robustness among all the $S_z$'s.

\section{Greedy Randomized Adaptive Search Procedure (GRASP) for \rls} 
\label{sec:graspp}

\subsection{Complexity analysis}
\label{sec:grasppcomp}
The size of subgraphs $|S|$ obtained during local search is $O(s)$.
Computing their top $t$ eigen-pairs takes $O(s^2t+st^2)$, where we use $e([S]) = O(s^2)$ as robust subgraphs are often  dense. 
To find the best improving node (L12), all nodes in the neighborhood ${\cal N}(S)\backslash S$ are evaluated, with worst-case size $O(n)$. As such, each expansion costs $O(ns^2t+nst^2)$.
With deletions incorporated (L3-4), the number of expansions can be arbitrarily large \cite{journals/jcss/JohnsonPY88},
however assuming $O(s)$ expansions are done, overall complexity becomes $O(ns^3t+ns^2t^2)$.
If all $t=|S|$ eigen-pairs are computed, the complexity is quadruple in $s$ and linear in $n$, which is feasible 
for small $s$. Otherwise, we exploit eigen-pair updates as in \greedymrs~to reduce computation. 


\vspace{-0.05in}
\subsection{Algorithm Variants}

Adapting \gmrs~for the \krls~can be easily done by returning the best $k$ 
(out of $T_{\max} \geq k$) distinct subgraphs  computed during the \grasp~iterations in Algorithm \ref{alg:grasp}. These subgraphs are likely to overlap, although one can incorporate constraints as to the extent of allowed overlap. 

For the \seedrls, we can initialize set $S$ with the seed nodes $U$ in construction (Alg. \ref{alg:cons} L1) while, during the local search phase, we never discard a node $u \in U$ from $S$ (Alg.\ref{alg:local} L4).

Finally, for the \rgs, we can waive the size constraint in the expansion step of local search (Alg.\ref{alg:local} L10).

\end{document}